\documentclass[reprint,aps,prx,twocolumn,groupedaddress, floatfix,nofootinbib]{revtex4-2}

\usepackage{graphicx}
\usepackage{bm}
\usepackage[utf8]{inputenc}
\usepackage{amsmath,amsfonts,calc}
\usepackage{comment}
\usepackage[normalem]{ulem}
\usepackage[usenames,dvipsnames]{xcolor}
\usepackage{graphicx}
\usepackage{scalerel}
\usepackage{mathtools}
\usepackage{oplotsymbl}
\usepackage[capitalise]{cleveref}
\usepackage{appendix}

\begin{document}
\title{\textcolor{black}{A driven quantum superconducting circuit with multiple tunable degeneracies}}
\author{Jayameenakshi Venkatraman}
\email{jaya.venkat@yale.edu,rodrigo.cortinas@yale.edu}
\thanks{these two authors contributed equally.}
\author{Rodrigo G. Corti\~nas}
\email{jaya.venkat@yale.edu,rodrigo.cortinas@yale.edu}
\thanks{these two authors contributed equally.}
\author{Nicholas E. Frattini}
\thanks{Present address: JILA, National Institute of Standards and Technology and the University of Colorado, Boulder, Colorado 80309, USA; Department of Physics, University of Colorado, Boulder, Colorado 80309, USA}
\author{Xu Xiao}
\author{Michel H. Devoret}
\email{michel.devoret@yale.edu}
\affiliation{Department of Applied Physics and Department of Physics, Yale University, New Haven, CT 06520, USA}
\date{\today}

\begin{abstract}
\color{black}
We present the experimental discovery of multiple simultaneous degeneracies in the spectrum of
a Kerr oscillator subjected to a squeezing drive. {This squeezing, in combination with the Kerr interaction creates an effective static two-well potential in the frame rotating at half the frequency of the sinusoidal driving force.} Remarkably, these degeneracies can
be turned on-and-off on demand, and their number is tunable. We find that when the detuning $\Delta$ between the
frequency of the oscillator and characteristic frequency of the drive equals an even multiple of
the Kerr coefficient $K$, $\Delta/K = 2m$, the oscillator displays $m + 1$ exact, parity-protected, spectral
degeneracies, insensitive to the drive amplitude. The degeneracies stem from the unusual destructive
interference of tunnel paths in the classically forbidden region of the double well static effective potential that models our experiment. Exploiting this interference, we measure a peaked enhancement of the incoherent well-switching lifetime creating a super-protected cat qubit in the ground state manifold of our oscillator. {Our results demonstrate the relationship between degeneracies and noise protection in quantum systems.}
\end{abstract}

\flushbottom
\maketitle
\thispagestyle{empty}

\color{black}
\textit{Introduction}  -- Degeneracies and their connection to symmetries play a pivotal role in physics. This connection leads to the emergence of noise-protected manifold of states for encoding and processing quantum information. For example, topological quantum systems exhibit global symmetries that result in degenerate ground states with inherent protection against local noise \cite{freedman2003}. To error-correct a quantum computation, the information must be protected by a symmetry such that the environment is blind to any unitary taking place within the manifold  of states \cite{ofek2016,google_surface_code}.

Atoms, like the hydrogen atom, exemplify the connection between symmetries and degeneracies through energy level degeneracies connected to spherical symmetry \cite{cohen1986,sakurai1995}. Superconducting circuits implement artificial atomic and molecular physics Hamiltonians with the virtue of in-situ tunability of parameters \cite{blais2021}. The pursuit of noise protection has led to the proposal and investigation of complex novel circuits, such as the $0-\pi$ qubit \cite{brooks2013}, whose near-degenerate qubit states are endowed with inherent resilience to decay and dephasing. However, the realization of such protected qubits often demands stringent circuit parameters.

In this work, we demonstrate an alternative approach to achieving circuit-level noise protection by employing a simple quantum nonlinear system, and engineering tunable spectral degeneracies within it. Our system consists of a sinusoidally-driven underdamped nonlinear oscillator. We realize a quantum double well with multiple degeneracies that can be turned on and off simply by varying the frequency of the drive. Specifically, when the detuning $\Delta$ between the
 frequency of the oscillator and the characteristic frequency of the drive equals an even multiple of the Kerr coefficient $K$, $\Delta/K = 2m$, the oscillator displays $m + 1$ exact, parity-protected, spectral degeneracies that are insensitive to the drive amplitude. Remarkably, these degeneracies correspond to the complete suppression of tunneling for excited states below the barrier in a double-well potential of finite height \cite{marthaler2007}.

Our experiment not only realizes for the first time, an elementary quantum optical system investigated theoretically \cite{Wielinga1993,marthaler2007,zhang2017} and uncovers new properties but also demonstrates new means to fight decoherence \cite{ruiz2022}. Specifically, we show that the quantum states at the bottom of the double well form a qubit manifold with a peaked inter-well transition lifetime, a phenomenon we name \textit{super-protection},  while remaining addressable. Thus, this qubit provides the basis for fault-tolerant syndrome measurement in quantum error correction \cite{Puri2019,grimsmo2021}.

\color{black}
\textit{Model system} -- We introduce our experimental system as a sinusoidally driven superconducting quantum circuit oscillator described by the time-dependent Hamiltonian
\begin{align}\label{eq:nl-osc-H}
\begin{split}
    \hat{\mathcal{H}}(t)/\hbar = \omega_o\hat a^\dagger \hat a &+ \frac{g_3}{3}(\hat a + \hat a^\dagger)^3 + \frac{g_4}{4}(\hat a + \hat a^\dagger)^4  \\
    &- i\Omega_d (\hat a - \hat a^\dagger)\cos\omega_d t,
\end{split}
\end{align}
where $\omega_o$ is the small oscillation frequency and $g_3, g_4 \ll \omega_o$ are the third and fourth-rank nonlinearities of the oscillator, $\hat a$ is the bosonic annihilation operator, and where the drive is specified by its amplitude $\Omega_d$ and frequency $\omega_d$. \Cref{eq:nl-osc-H} models a SNAIL transmon that is charge-driven at frequency $\omega_d$. It is the electrical circuit analog of an asymmetric mechanical pendulum capable of three and four-wave mixing \cite{frattini2021}. The experimental setup has been described in \cite{frattini2022}. The drive is configured so that its second subharmonic $\omega_d/2$ lies in the vicinity of the SNAIL transmon  resonance at $\omega_a = \omega_o + 3 g_4 - 20 g_3^2/3 \omega_o + \mathcal{O}(g_3^3/\omega_o^2)$.

\begin{figure*}
    \includegraphics[width = 0.99\textwidth]{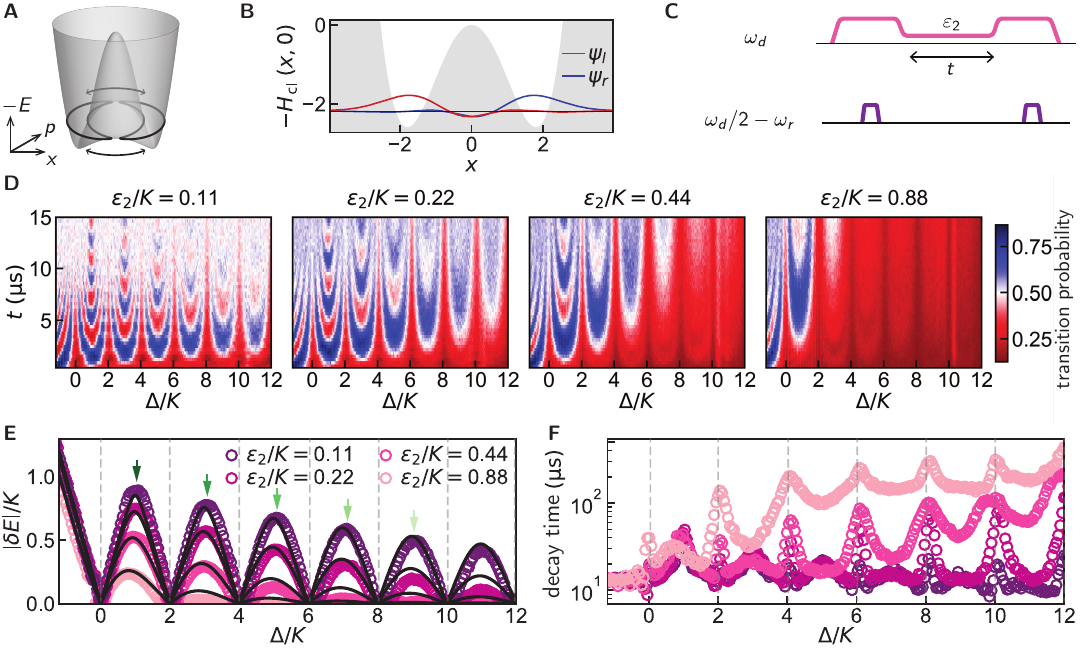}
    \caption{\textbf{Tunnel-driven Rabi oscillations in the ground state manifold and their periodic cancellation}. \textbf{A} Energy surface associated with \cref{eq:Heff} in the classical limit for $\Delta/K = 3$ and $\epsilon_2/K = 0.11$. The orbits shown with black lines are obtained by semiclassical action quantization and represent the ground states (see supplement). Bidirectional arrows represent the two interfering WKB tunneling paths. \textbf{B} Cut of the energy surface in \textbf{A} at $p = 0$ (see supplement). The classically forbidden region is marked in grey. The left and right localized wavefunctions are indicated in red and blue. \textbf{C} Pulse sequence for \textbf{D}. The pink line represents the squeezing drive at frequency $\omega_d$ and the purple lines represent the preparation and readout drives at frequency $\omega_{d}/2 - \omega_r$. \textbf{D} Time-domain Rabi oscillation measurement of inter-well tunneling probability (color) as a function of $\Delta^{\mathrm{bare}}$, taken here as $\Delta$ (see text), for $\epsilon_2/K =$ 0.11, 0.22, 0.44, and 0.88. The extracted tunneling amplitudes from \textbf{D} are shown as open circles in \textbf{E}. The black lines in \textbf{E} correspond to the transition energy between the lowest eigenstates obtained from an exact diagonalization of \cref{eq:Heff}. A comparison of the extracted tunneling rate with a semiclassical WKB calculation is presented in the supplement. Green arrows in $\textbf{E}$ denote the condition for constructive interference of tunneling and correspond to the measurements shown in \Cref{fig:2}. We extract the value of the Kerr coefficient $K$ from this data and note that it is consistent, within experimental inaccuracies, with an independent saturation spectroscopy measurement of the Fock qubit in the absence of the squeezing drive (see supplement). \textbf{F} Decay time of the tunnel-driven Rabi oscillations for different values of $\Delta$ and $\epsilon_2$ in \textbf{D}. Sharp peaks in the decay time are clearly visible for $\Delta/K = 2m$, $m$ being a non-negative integer.}
    \label{fig:1}
\end{figure*}

Under an approximation beyond the rotating-wave that captures the averaged behaviour of this rapidly driven nonlinear superconducting circuit, the dynamics governed by \cref{eq:nl-osc-H} is well-described by the static effective Hamiltonian \cite{venkatraman2022,grimm2020,frattini2022}
\begin{align}
\label{eq:Heff}
    \hat{H}/\hbar = \Delta \hat a ^\dagger \hat a - K \hat{a}^{\dagger 2} \hat{a}^2 + \epsilon_2 ( \hat{a}^{\dagger 2} + \hat{a}^{2}).
\end{align}
\Cref{eq:Heff} corresponds to an elementary quantum system: a Kerr oscillator dressed by a squeeze-drive. In  \cref{eq:Heff}, the detuning parameter is given by \\ $\Delta = \Delta^{\mathrm{bare}} + \delta^{\mathrm{ac}}$, where $\Delta^{\mathrm{bare}} = \omega_a - \omega_d/2$, with $|\Delta^{\mathrm{bare}}| \ll \omega_a$, and where $\delta^{\mathrm{ac}}$ corresponds to the ac Stark shift: $\delta^{\mathrm{ac}} = (6 g_4 - 9 g_3^2/ \omega_o  + \mathcal{O}(g_3^3/\omega_o^2)) |\Pi|^2$, where $|\Pi| = \Omega_d \omega_d/(\omega_d^2 - \omega_o^2)$. For our system, we measure $\omega_a/2\pi = 6.035~\mathrm{GHz}$. The Kerr coefficient arises from the bare $g_3$ and $g_4$ nonlinearities of the circuit, which are themselves controlled in situ by a magnetic field, and is given by $K =  10 g_3^2/3 \omega_o - 3 g_4/2 + \mathcal{O}(g_3^3/\omega_o^2))$. In our experiment, we measure it to be $K/2\pi= 316.8~\mathrm{kHz}$ (see \cref{fig:1}). A crucial component of the experiment, the squeezing drive amplitude $\epsilon_2 = g_3 |\Pi| $, is generated by the near-resonant three-wave mixing process between one drive excitation and two oscillator excitations.  Due to the relatively small $K$ compared to a standard transmon \cite{blais2021}, our experiment has a negligible ac Stark shifts for $\epsilon_2/K \lesssim 1$, so that in this regime $\delta^{\mathrm{ac}}/ K \lesssim 1\%$. Therefore, in this regime, $\Delta$ can be approximated by $\Delta^{\mathrm{bare}} = \omega_a - \omega_d/2$. By taking $K$ to provide the natural units for our system, the Hamiltonian is completely determined by only two dimensionless parameters: $\Delta/K$ and $\epsilon_2/K$, where the former is controlled by the drive frequency and the latter is directly proportional to the drive amplitude. We thus have independent real-time control of all relevant Hamiltonian parameters. Lastly, in our experiment, the single-photon lifetime of the undriven SNAIL transmon is $T_1 = 20$~\textmu s and the Ramsey coherence between its lowest-lying eigenstates is $T_{2 \mathrm{R}} = 3.8$~\textmu s.

\textit{Experiment and results} -- We first experimentally demonstrate the cancellation of tunneling in the ground state manifold. In \Cref{fig:1}\textbf{A}, we show the classical limit of the energy surface associated with \cref{eq:Heff} for $\Delta/K = 3, \epsilon_2/K = 0.11$, as a function of phase-space coordinates. The arrows indicate the two WKB tunneling paths \cite{marthaler2006}. Furthermore, we show in \Cref{fig:1}\textbf{B}, the  wavefunctions corresponding to the ground state manifold. Note that these are not the energy eigenstates but their even and odd superpositions, which are localized in the left and right wells. Importantly, in the classically forbidden region, marked in grey, oscillations accompany the expected decay of the wavefunctions \cite{marthaler2007}. To observe coherent cancellation of tunneling in the ground state manifold, we prepare a localized well state and measure its tunneling probability as a function of time for different values of $\Delta$ and $\epsilon_2$. We present the measurement protocol in  \Cref{fig:1}\textbf{C}. The preparation is done by rapidly turning on the squeezing drive until an amplitude of $\epsilon_2/K = 8.7$ is reached. We subsequently wait for $5 T_1$ for the system to relax to its steady state in the presence of the squeezing drive and measure, in a quantum nondemolition (QND) manner, the quadrature containing the which-well information. This measurement projects the system into one of the wells. It is done by the microwave activation of a parametric beam splitter interaction between the squeeze-driven Kerr oscillator and a readout resonator strongly coupled to a quantum-limited amplifier chain. We refer the reader to \cite{frattini2022} for experimental details, where the preparation-by-measurement procedure for our system was introduced. This readout protocol yields a stabilized fluorescence signal revealing the quadrature measurement outcome, while the squeezing drive sustains the circuit oscillation. After the preparation, we adiabatically lower the squeezing drive amplitude in a duration $1.6~\mathrm{\mu s}$ $\gtrsim \pi/K$.\footnote{Note that this adiabaticity condition pertains to the gap between the ground and first excited pair of states. We do not need to be adiabatic with respect to the two tunnel split states within the ground state manifold since they have opposite parity and the parity preserving squeezing drive will not couple them.} The depth of the wells, which increases with $\epsilon_2/K$ (see supplement), is then reduced so that the tunnel effect becomes observable.  We then wait for a variable amount of time before adiabatically raising the squeezing drive amplitude to its initial value. Finally, we measure which well the system has adopted. 

The data for this tunneling measurement is shown in \Cref{fig:1}\textbf{D}, where we interpret the oscillating color pattern as tunnel-driven Rabi oscillations. The periodic cancellation of tunneling at $\Delta/K=2m$, where $m$ is a non-negative integer, is clearly visible as a divergence of the Rabi period. We extract the tunneling amplitude $|\delta E|$ from our data by fitting the oscillation frequency with an exponentially decaying sinusoid and plot this frequency in \Cref{fig:1}\textbf{E}, where the data-point color corresponds to the value of $\epsilon_2$ (see supplement for calibration of $\epsilon_2$). The black lines,  obtained from an exact diagonalization  of the static effective Hamiltonian \cref{eq:Heff}, correspond to the energy difference between levels in the ground state manifold. The cancellation of tunneling for the ground state manifold in a parametrically modulated oscillator was predicted by \cite{marthaler2007} where, using a semiclassical WKB method, the authors found that this multi-path interference effect is due to, and accompanied by, oscillations of the wavefunction crossing zero in the classically forbidden region. Here, we find good agreement between our experiment and their WKB prediction (see supplement). Note that, across the zero of the tunneling amplitude, the bonding and anti-bonding superposition of well states alternate as the ground state. Specifically, for $\Delta/K = 4m+1$, the ground state is the bonding superposition of well states (see supplement). In \cref{fig:1}\textbf{F}, we further plot the extracted decay time of the tunneling oscillations as a function of $\Delta$, and find sharp peaks when $\Delta/K = 2m$, besides an overall continuous increase of the decay time with $\Delta$ and $\epsilon_2$. The peaks at $\Delta/K = 2m$ arise from the degeneracies in the excited state spectrum at this condition and are discussed later in the text.

Importantly, the dynamics of the two-level system in \Cref{fig:1}\textbf{D} suggest a new type of bosonic encoding of information that we call the $\Delta$-Kerr-cat qubit. The north and south poles of the corresponding Bloch sphere, a generalization of the $\Delta = 0$ one \cite{puri2017,grimm2020,frattini2022}, is defined by the cat states formed by the lowest pair of eigenstates of \cref{eq:Heff}. In this picture, a tunnel-Rabi cycle in \Cref{fig:1}\textbf{D} for a fixed $\Delta/K \ne 2m$ corresponds to a travel along the equator. For $\Delta/K = 2m$, this travel is prohibited. Note that when $\Delta/K = 2m+1$, the tunneling amplitude is maximum and is first-order insensitive to fluctuations of $\Delta$.

\begin{figure}[t!]
\includegraphics{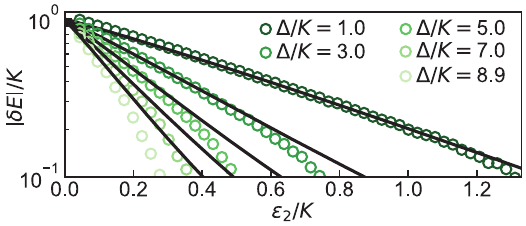}
\caption{\textbf{Exponential reduction of tunnel splitting as a function of $\epsilon_2$ in the ground state manifold.} Extracted tunnel splitting (open circles) for the first five local maxima in \Cref{fig:1}\textbf{E} as marked by the color coded arrows. Experimental sequence as in \Cref{fig:1}\textbf{E}. For the raw color data, see Figure 3 in the supplement. Black lines are obtained from a Hamiltonian diagonalization of \cref{eq:Heff} with no adjustable parameters. For comparison with a semiclassical WKB calculation, see supplement. Note that for small tunneling amplitude, dissipation plays a relevant role and the Hamiltonian model used here is insufficient.}
\label{fig:2}
\end{figure}

From \Cref{fig:1}\textbf{E}, we also see that, besides the \textit{discrete cancellation} of tunneling  at $\Delta/K = 2m$, tunneling in the ground state manifold is overall \textit{continuously reduced} with both $\Delta$ and $\epsilon_2$. This reflects the well-known symmetry of the double well, which is broken by tunnel coupling. The approximate symmetry is restored with increasing $\Delta$ and $\epsilon_2$ because both parameters explicitly control the barrier height and thus exponentially control the tunneling amplitude $|\delta E|$. Theory predicts that the larger the detuning $\Delta$, the faster the tunneling reduction with the squeezing drive amplitude $\epsilon_2$ (see supplement). We have measured this effect by measuring the tunneling amplitude as a function of $\epsilon_2$ for different constructive tunneling conditions corresponding to $\Delta/K = 2m+1$. The data is presented in \Cref{fig:2}. The exponential insensitivity, around $\Delta = 0$, to fluctuations of $\Delta$ due to a noisy $\omega_a$, as a function of $\epsilon_2$, was predicted by \cite{puri2017}  and thus proposed as a resource for quantum information. This insensitivity was a key motivation for realizing the Kerr-cat qubit experimentally \cite{grimm2020}. The insensitivity of the ground state manifold to detuning as a function of $\epsilon_2$ is directly observed here for the first time. Note from \Cref{fig:1}\textbf{E} that for $\Delta < 0 $, in the parameter regime $\epsilon_2/K < 1$, the tunneling amplitude $|\delta E|$ is weakly dependent on $\epsilon_2$, whereas for $\Delta > 0$, it is strongly dependent on $\epsilon_2$. This weak dependence for $\Delta<0$ is expected since the barrier height vanishes for small values of $\epsilon_2/K$.\footnote{
In the absence of dissipation, the metapotential acquires two wells as soon as $\epsilon_2,\, \Delta > 0$, i.e. there is no threshold for bifurcation of the driven oscillator.  In our quantum experiment, this threshold is finite but is, relatively speaking, extremely small since and is set by $\epsilon_2^2 > (\Delta^2 + T_1^{-2}/4)/4$ (see \cite{frattini2021}).} Our finding shows that new operating points at even, positive values of $\Delta/K$ will increase the resilience of ground-state qubit encoding to detuning-like noise.

\begin{figure*}
\includegraphics[width = 0.99\textwidth]{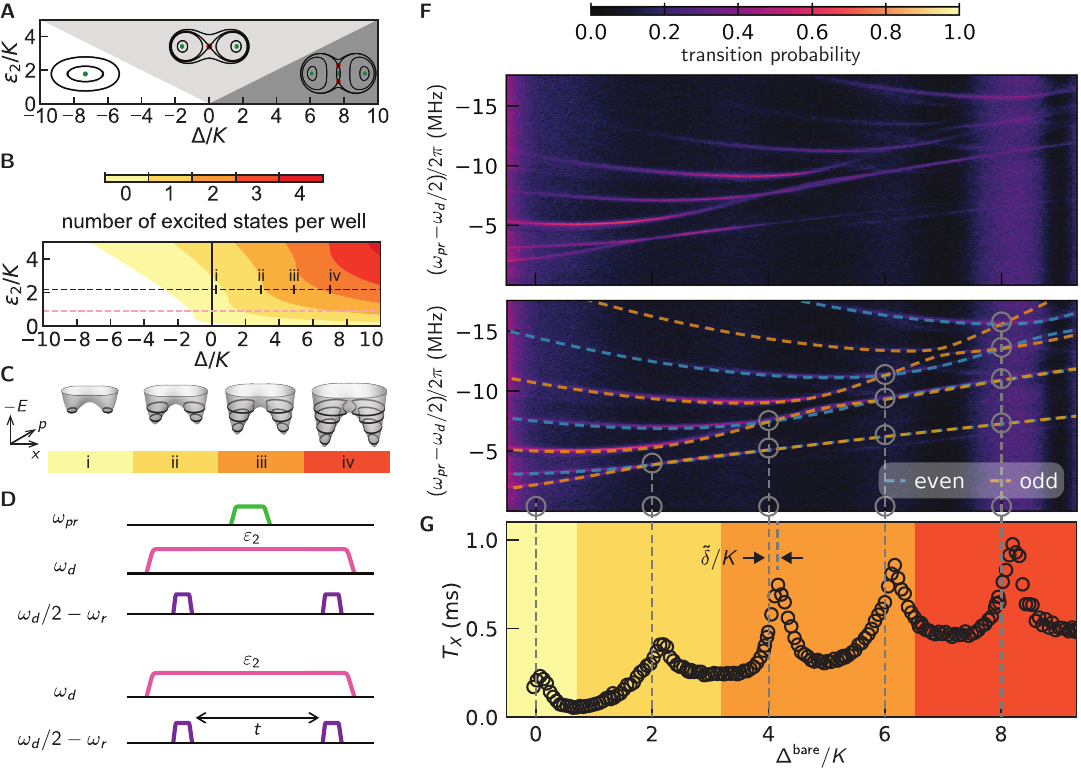}
\caption{\textbf{Spectroscopic measurements of coherent and periodic cancellation of tunnel splitting in the excited state spectrum}.
\textbf{A} Classical phase diagram for the Kerr oscillator with parametric squeezing, also called the period-doubling bifurcation diagram. \textbf{B} Quantum phase diagram to count in-well excited states. White lines separate single-node, double-node, and triple-node phases. Colors represent contours of constant action on the energy surface associated with \cref{eq:Heff}. Dashed pink line corresponds to $\epsilon_2/K = 0.88$, the maximum value of squeezing drive amplitude in \cref{fig:1}. Dashed black line corresponds to $\epsilon_2/K = 2.17$, the value of squeezing drive amplitude used in Figs. 3\textbf{F} and 3\textbf{G}. Energy surfaces for $\epsilon_2/K = 2.17$ and (i) $\Delta/K = 0.5$, (ii) $\Delta/K = 3$, (iii) $\Delta/K = 5$, and (iv) $\Delta/K = 7$. Bohr-like obits are indicated as black curves (see supplement for more details). \textbf{D} Pulse sequence for \textbf{F}. The green line represents the weak spectroscopic probe tone at frequency $\omega_{{pr}}$. The pink line represents the squeezing drive at frequency $\omega_d$ and the purple lines represent the preparation and readout drives at frequency $\omega_{d}/2 - \omega_r$. \textbf{E} Pulse sequence for \textbf{G}. \textbf{F} (upper panel) Frequency-domain measurement of well-transition probability (color) via excited states as a function of $\Delta$ for $\epsilon_2/K = 2.17$. The power of the perturbative spectroscopic probe is increased as $\omega_{pr}$ is decreased to compensate for the lower matrix element connecting the ground state with the higher excited levels, yet is kept weak enough to preserve the parity conservation rules of \cref{eq:Heff}. \textbf{F} (lower panel) Dashed lines plotted on top of experimental data (same as in upper panel) correspond to transition energies obtained by performing an exact diagonalization of \cref{eq:Heff} with no adjustable parameters. The Kerr coefficient is calibrated via time-domain measurements in \Cref{fig:1}\textbf{E}.  \textbf{G} Measured well-switching time under incoherent environmental-induced evolution as a function $\Delta$ for $\epsilon_2/K \approx 2.17$. Background color in \textbf{G} marks the number of excited states per well following semiclassical orbit quantization.}
\label{fig:3}
\end{figure*}

\textcolor{black}{Moving to the pairs of excited states above the ground state manifold, do they also present observable degeneracies as a function of $\Delta/K$?} In order to deepen our understanding of this problem, we first examine the classical energy surface associated with \cref{eq:Heff} via the period doubling phase diagram \cite{wustmann2019} shown in \Cref{fig:3}\textbf{A}.  In the classical limit (see supplement), the parameter space spanned by $\Delta/K$ and $\epsilon_2/K$ is divided by two phase transitions located at $\Delta =\pm 2 \epsilon_2$. The different phases are characterized by the number of stable nodes (attractors) in the classical metapotential and we refer to them as the single-, double-, and \linebreak  triple-node phases. These phases correspond to different metapotential topologies. We show them as contour line insets in \Cref{fig:3}\textbf{A}, representing classical orbits. The single-node phase occurs for $\Delta <-2 \epsilon_2$, and presents only one well. For $\Delta \ge -2 \epsilon_2$, the oscillator has bifurcated and the classical metapotential acquires two wells. In the presence of dissipation, these wells house stable nodes. The emergent ground state manifold has been exploited, for $\Delta = 0$, in the Kerr-cat qubit \cite{grimm2020,frattini2022}. In the interval $-2\epsilon_2 \le \Delta < 2 \epsilon_2$, an unstable extremum (saddle point) appears at the origin. For $\Delta \ge 2 \epsilon_2$, the saddle point at the origin splits into two saddle points and an attractor reappears at the origin. The barrier height of the classical metapotential is given by $(\Delta + 2\epsilon_2)^2/4K$ in the double-node phase and by $2\epsilon_2 \Delta/K$ in the triple-node phase (see supplement). To count the number of excited states that have sunk under the barrier, we further introduce in \Cref{fig:3}\textbf{B} a semi-classical phase diagram of the squeeze-driven Kerr oscillator. Following the Einstein-Brillouin-Keller method, which generalizes the notion of Bohr orbits, we quantize the action enclosed in the metapotential well below the height of the barrier and obtain the number of in-well excited states. In \Cref{fig:3}\textbf{C}, we present the corresponding orbits in the energy surface for a fixed value of $\epsilon_2/K = 2.17$ and four values of $\Delta/K$. We validate this simple, semiclassical picture with a fully quantum mechanical calculation of the Wigner functions of localized states in the ground and excited state manifold (see supplement). It is clear from this analysis that, by increasing $\epsilon_2$ and $\Delta$, and therefore the barrier height, not only the ground state manifold but even the excited state manifolds become progressively ensconced in the wells, and we thus expect the tunneling between the wells to be drastically reduced. 

Besides the overall \textit{continuous reduction} of tunneling, the excited state manifold of the squeeze-driven Kerr oscillator experiences a \textit{discrete cancellation} of tunneling when $\Delta/K = 2m$. Since the squeezing interaction preserves photon parity, levels belonging to the even and odd sector of the Kerr Hamiltonian remain decoupled and repeatedly cross at values of $\Delta/K$ corresponding to even integers. This braiding induces $m+1$ perfect degeneracies at $\Delta/K = 2m$. Moreover, the corresponding eigenstates have a closed-form expression in the Fock basis. Remarkably, these features are independent of the value of $\epsilon_2$, reflecting a particular, unappreciated symmetry of our Hamiltonian \cref{eq:Heff} (see supplement).

\begin{figure}[t!]
\includegraphics{{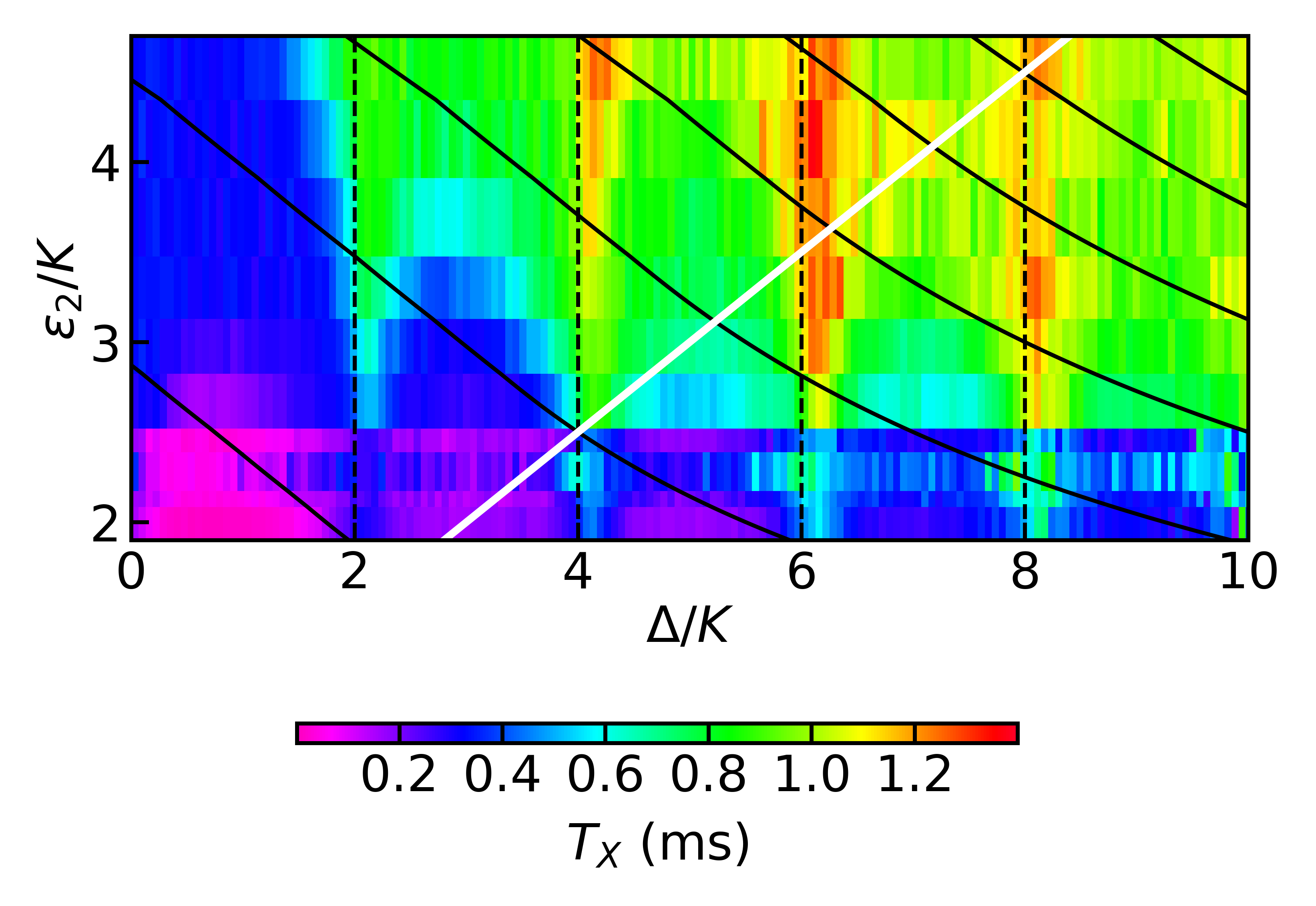}}
\caption{\textbf{Color plot of $T_X$ as a function of $\Delta/K$ and $\epsilon_2/K$}. White line marks the transition from a two-node to a three-node metapotential. Black solid lines mark contours of constant barrier height. Increasing both $\Delta/K$ and $\epsilon_2/K$ yields fastest enhancement in $T_X$ as predicted by \Cref{fig:3}\textbf{B}. The additional enhancement by the coherent cancellation of excited state tunneling at $\Delta/K = 2m$ stands out. The pulse sequence for the measurement is shown in \Cref{fig:3}\textbf{E}.}
\label{fig:2D-lifetimes}
\end{figure}

Both the \textit{discrete cancellation} and the overall \textit{continuous reduction} of tunneling now in the excited state manifold of the squeeze-driven Kerr oscillator is accessed by performing spectroscopy measurements as a function of $\Delta$, which we show in \Cref{fig:3}\textbf{F} for $\epsilon_2/K = 2.17$. The measurement protocol is shown in \Cref{fig:3}\textbf{D}. We prepare a localized well state in a manner that is similar to the protocols of \Cref{fig:1,fig:2}. To locate the frequency of the excited states, we apply a probe tone at variable frequency in the vicinity of the SNAIL transmon resonance $\omega_a$ and measure the well-switching probability. When the probe is resonant with a transition to a state close to the barrier maximum, this probability is increased. The experimental results are shown in \Cref{fig:3}\textbf{F}. The colored dashed lines (orange and blue) in the lower panel are obtained from an exact diagonalization of the static effective Hamiltonian \cref{eq:Heff} with no adjustable parameters. The crossings of levels are marked with circles. The data also shows that the level crossings are accompanied by a continuous reduction of the braiding amplitude with $\Delta$. The corresponding reduction of the tunnel splitting is the manifestation associated with a generic double-well Hamiltonian while the braiding reflects interference specific to our particular Hamiltonian, resulting from its underlying driven character. The level of experimental control achieved allows us to observe in this data the joint presence of the \textit{exact discrete symmetry} and the \textit{approximate continuous symmetry} in our bosonic system.

An important consequence of the cancellation of tunneling in the excited state spectrum is the periodic enhancement of the well-switching time under incoherent environment-induced evolution. This time scale corresponds to the transverse relaxation time, $T_X$, of a new bosonic qubit: a $\Delta$-variant of the Kerr-cat qubit \cite{puri2017,frattini2021} as mentioned earlier. To measure $T_X$, we prepare a localized well state by measurement, and wait for a variable amount of time before measuring the which-well information. We show the pulse sequence in \Cref{fig:3}\textbf{E}. We obtain $T_X$ by fitting a decaying exponential function to the measured well-transition probability for each value of $\Delta$ and plot the result in \Cref{fig:3}\textbf{G}. Note that we have chosen the squeezing drive amplitude, identical to that of \Cref{fig:3}\textbf{F}, as $\epsilon_2/K = 2.17$. Around values of $\Delta/K$ corresponding to even integers, the variation of $T_X$ presents sharp peaks. The location of the peaks corresponds to the degeneracy condition in the excited state spectrum, associated with coherent cancellation of tunneling and the blocking of noise-induced well-switching pathways via the excited states. The systematic right-offset $\tilde{\delta}/K$ of each peak from an even integer, is 15\%. About 5\% can be attributed to the ac Stark shift $\delta^{\mathrm{ac}}$ for this photon number, given the accuracy of our knowledge of the experimental parameters. We do not have an explanation for the remaining 10\%, but we suspect it could be explained by higher-order terms in our static effective Hamiltonian. Note that this explanation is still compatible with the perfect alignment of the cancellation points with even integers in \Cref{fig:1}\textbf{F} for $\epsilon_2/K < 1$, since for that case the ac Stark shift is negligible. Note also that this offset could provide access, within experimental accuracy, via the ac Stark shift, to the nonlinear parameters of \cref{eq:nl-osc-H}. 

The data in \Cref{fig:3}\textbf{G} also shows that the discrete peaks are accompanied by a monotonic baseline increase, a direct manifestation of the overall continuous tunneling reduction in the spectrum versus $\Delta$. The background colored stripes represent the number of in-well excited states found via the action quantization method discussed above and in the supplement. Continuing with this semiclassical picture, we interpret the slowdown in the growth of $T_X$ for $\Delta/K \gtrsim 5$ as the slowdown in the growth of the barrier height as one crosses over from the double-node, where the barrier height $\propto (\Delta + 2\epsilon_2)^2$, to the triple-node phase, where the barrier height $\propto \Delta \epsilon_2$. Indeed, this is the quantum manifestation of the classical phase transition from the double-node to the triple-node phase.

Thus, whether the theoretical framework is classical, semiclassical, or quantum, the predicted $T_X$ will increase with both $\epsilon_2$ and $\Delta$. While $\epsilon_2$ and $\Delta$ contribute via the overall continuous reduction of tunneling \cite{frattini2022}, only $\Delta$ controls the discrete cancellation of tunneling. We verify this prediction by measuring $T_X$ while varying simultaneously both Hamiltonian parameters. We present the result of this experiment in \Cref{fig:2D-lifetimes}. We further plot contours of constant barrier height in black, and the expected separation between the double-node and triple-node metapotential as a white line. The system lying deeply in the quantum regime, we do not expect any sharp features along this line. As expected, following the gradient of the barrier height, one observes the fastest gain in $T_X$, with a maximum of $T_X = 1.3~\mathrm{ms}$ for $\Delta/K = 6$ and $\epsilon_2/K = 4$. Increasing the lifetime by increasing $\epsilon_2$ presents limitations, since strong drives are known to cause undesired effects in driven nonlinear systems (see \cite{xiao2022,venkat_EffLindbladian2022} and supplement).

One could argue that $\Delta = 0$ provides an important factorization condition that guarantees that the ground state manifold is spanned by exact coherent states (see \cite{puri2017} and supplement). Indeed, this is an asset for quantum information, since these states are eigenstates of the single-photon loss operator $\hat a$ \cite{mirrahimi2014}. However, this desirable property is traded for the advantages discussed earlier when $\Delta/K =2m$, $m\geq 1$. Even if the $\Delta$-variant of the Kerr-cat qubit suffers from quantum heating and quantum diffusion \cite{marthaler2006,dykman2012,ong2013} at zero temperature resulting from the squeezed nature of its ground states, these effects are small and, as we show in the experiments reported here and in \cite{frattini2022}, the well-states of the Kerr-cat live longer than its $\Delta = 0$ parent, even at finite temperature.

\textit{Discussion} -- Although quantum tunneling was discovered nearly a century ago \cite{Merzbacher2002} and observed since in a variety of natural and synthetic systems, the treatment of tunneling is usually limited to the ground states of the system and has rarely been discussed for excited states in the literature, as we elaborate in the following survey. The phenomenology of ground state tunneling has been studied in cold atoms \cite{ramos2020} in three-dimensional optical lattices \cite{folling2007}, optical tweezers \cite{Kaufman2014}, ion traps \cite{noguchi2014} and in quantum dots \cite{Hsiao2020}. In Josephson tunnel circuits, quantum tunneling of the phase variable was first observed by Devoret, Martinis, and Clarke \cite{devoret1985} and since then exploited in several other experiments \cite{vijay2009}. Furthermore, the tunnel effect has been involved in quantum simulation \cite{Bloch2012}, in Floquet engineering of topological phases of matter and to generate artificial gauge fields with no static analog \cite{goldman2014,wintersperger2020}. The quantum interference of tunneling for the ground states of a large spin system was measured previously in a cluster of eight iron atoms by Wernsdorfer and Sessoli \cite{wernsdorfer1999} (see also \cite{alexandradinata2018}).

Weilinga and Milburn \cite{Wielinga1993} first identified that the quantum optical model in \cref{eq:Heff} exhibits ground state tunneling for a particular value of $\Delta$. Marthaler and Dykman \cite{marthaler2006,marthaler2007} developed a WKB treatment for a range of the $\Delta$ parameter, and predicted that, for this model, the tunnel splitting of the ground state manifold crosses zero periodically and is accompanied by oscillation of the wavefunction in the classically forbidden region.

Our work is the first experimental realization of the longstanding theoretical proposals of the last paragraph. It is similar, but different, to the phenomenology of the ``coherent destruction of tunneling'', discovered theoretically by Grossmann et al. \cite{grossmann1991} and observed experimentally in cold atoms \cite{lignier2007,chen2011}. Indeed, the dynamical tunneling in our experiment is in sharp contrast with photon-assisted or suppressed tunneling in weakly driven double-well potentials. Firstly, our tunneling is completely dynamical, i.e., the tunneling barrier vanishes in the absence of the drive. Secondly, and most importantly, our work extends the coherent cancellation of tunneling to all the excited states in the well. The periodic resonance condition $\Delta/K = 2m$, shared for the $m+1$ first pairs of excited levels, is independent of the drive amplitude. Remarkably, under this multi-state resonance condition, the first  $2(m+1)$ oscillator states have a closed-form expression in the Fock basis (see supplement). We further emphasize that the dynamical tunneling in our work is distinct from chaos-assisted dynamical tunneling \cite{tomsovic1994} observations made in ultracold atoms over three decades ago \cite{hensinger2001,tomsovic1994}; remarkably our strongly driven nonlinear system remains integrable. To the best of our knowledge, our work corresponds to the discovery and the first demonstration of the exact simultaneous cancellation of the tunnel splitting for the ground and excited states. Our data featuring the incoherent dynamics can be qualitatively modeled by a Lindbladian treatment that we present in the supplement, yet more research on the decoherence of driven nonlinear driven systems is needed to get a quantitative agreement (see \cite{venkat_EffLindbladian2022}).

As a resource for quantum information, the squeeze-driven Kerr oscillator for $\Delta = 0$, was identified in theory proposals by Cochrane, Milburn, and Munro \cite{cochrane1999} and Puri, Boutin, and Blais \cite{puri2017} due to its exponential resilience to low frequency noise and was proposed for a bosonic code. The code was implemented for the first time in circuits \cite{grimm2020}. 
Bistability for non-zero $\Delta$ was predicted by Roberts and Clerk in \cite{roberts2020}. Our work demonstrates this bistability experimentally through the lifetime peaks in \cref{fig:3}\textbf{G} and explains the peaks as a fingerprint of the observed spectral degeneracies in \cref{fig:3}\textbf{F}.
Furthermore, the resilience to noise in the non-zero $\Delta$ case is demonstrated through \cref{fig:1}\textbf{E} and \cref{fig:2}.

\color{black}\textit{Conclusion} -- We have observed multiple degeneracies between pairs of states in a quantum double-well system, resulting from the interplay of quantum tunneling and quantum interference. Our work showcases the tunability of these degeneracies in number and the ability to rapidly activate or deactivate them. Furthermore, we have identified the drive frequency as a critical control parameter, governing not only a discrete exact symmetry in \cref{eq:Heff}, manifested as exact degeneracies, but also a continuous approximate symmetry that leads to an overall exponential reduction of tunnel splitting in both ground and excited states of our oscillator. This high degree of quantum control culminates in a significant reduction of incoherent well-flip dynamics, enabling the creation of a super-protected cat-qubit—the $\Delta$-Kerr-cat qubit. Our demonstration of the continuous Z-gate \cite{kanao2022,Puri2019} adds valuable capability to the single qubit gate-set for cat qubits, offering new tools for quantum computation \cite{mirrahimi2014,kanao2022,puri2017,grimm2020,frattini2022,raimond2006,puri2017_ising,puri2017_2}. With comprehensive control over the parameter space of an individual squeeze-driven Kerr oscillator and the ability to measure its spectrum as a function of these parameters, our system holds immense significance in the theories of Quantum Phase Transitions (QPT) \cite{iachello1995}, Excited State QPT (ESQPT) \cite{chavez2022}, and Dissipative QPT (DQPT) \cite{gravina2022}. Incidentally, the phase portrait presented in \cref{fig:3}\textbf{A} is very similar to the one in Figure 1 in \cite{nader2021}. Moreover, our findings underscore the potential of superconducting circuits for simulating symmetries, providing the unprecedented advantage of in situ tunability. This breakthrough opens up new research avenues in the simulation of atomic, molecular, and nuclear physics.
\color{black}

A quasi-spin symmetry and an algebraic structure underlying the measured degeneracies were recently discovered by F. Iachello and will be discussed in a forthcoming paper.

After our experiments were performed, we learned that the degeneracies in our squeeze-driven Kerr oscillator were studied theoretically by our colleagues in the QUANTIC group in INRIA, Paris \cite{ruiz2022}.

\textit{Acknowledgements} -- We acknowledge Vladislav Kurilovich for pointing to us the peculiarity of the amplitude independence of the multi-level degeneracies in our model. We thank Charlotte G. L. Bøttcher, Steven M. Girvin, Leonid Glazman, Francesco Iachello, Alessandro Miano, Shruti Puri, and Qile Su for useful discussions. R. G. C acknowledges useful discussions with Lea Santos, Mazyar Mirrahimi, Diego Ruiz, and Jérémie Guillaud.
This research was sponsored by the Air Force Office of Scientific Research under award number FA9550-19-1-0399, by the Army Research Office (ARO) under grant numbers W911NF-18-1-0212 and W911NF-16-1-0349, and by the National Science Foundation (NSF) under award numbers 1941583 and 2124511, and by the U.S. Department of  Energy, Office of Science, National Quantum Information Science Research Centers, Co-design Center for Quantum Advantage (C2QA) under contract number DE-SC0012704. The views and conclusions contained in this document are those of the authors and should not be interpreted as representing the official policies, either expressed or implied, of the U.S. Government. The U.S. Government is authorized to reproduce and distribute reprints for Government purposes notwithstanding any copyright notation herein. Fabrication facilities use was supported by the Yale Institute for Nanoscience and Quantum Engineering (YINQE) and the Yale SEAS Cleanroom.

\textit{Author contributions} -- N.E.F. designed the sample package and fabricated the device. N.E.F. and R.G.C. built the measurement setup. X.X. and J.V. identified $\Delta$ as a useful knob and J.V. proposed its experimental implementation. J.V. and R.G.C. designed and performed experiments and analyzed data. J.V. and R.G.C. performed theoretical calculations and simulations. M.H.D., X.X., and N.E.F. discussed the project with J.V. and R.G.C. and provided insight. J.V., R.G.C., and M.H.D. wrote the manuscript with input from all the authors.
\bibliography{bib_main}

\end{document}


\title{\textcolor{black}{A driven quantum superconducting circuit with multiple tunable degeneracies}}
\author{Jayameenakshi Venkatraman}
\email{jaya.venkat@yale.edu,rodrigo.cortinas@yale.edu}
\thanks{these two authors contributed equally.}
\author{Rodrigo G. Corti\~nas}
\email{jaya.venkat@yale.edu,rodrigo.cortinas@yale.edu}
\thanks{these two authors contributed equally.}
\author{Nicholas E. Frattini}
\thanks{Present address: JILA, National Institute of Standards and Technology and the University of Colorado, Boulder, Colorado, 80309, USA; Department of Physics, University of Colorado, Boulder, Colorado, 80309, USA}
\author{Xu Xiao}
\author{Michel H. Devoret}
\email{michel.devoret@yale.edu}
\affiliation{Department of Applied Physics and Department of Physics, Yale University, New Haven, CT 06520, USA}
\date{\today}

\flushbottom
\maketitle

The Supplemental text is organized as follows. In \cref{sec:e2-calib}, we detail the experimental calibration of Hamiltonian parameters in Eq. (2) of the main text; specifically, in \cref{sec:e2-calib} we present a calibration of the squeezing drive $\epsilon_2$ and in \cref{sec:K-calib}, we present a measurement of the Kerr coefficient $K$. In \cref{sec:exp-red-data,sec:lifetime}, we present further experimental results supplementing Figures 2 and 4 in the main text. 

In \cref{sec:notation} and the following sections, we switch gears and detail our theoretical models. First, in \cref{sec:notation} we formally introduce the notation employed throughout this work. We then present in \cref{sec:relation} two well-known squeeze-driven Kerr oscillator Hamiltonians that were introduced in the literature and comment on the relationships between them. In \cref{sec:rep-H-H}, we introduce the operator and phase space formulation of our particular squeeze-driven Kerr oscillator effective Hamiltonian and discuss its classical limit. 

In \Cref{sec:wf_lowest,sec:semi,sec:spectrum}, we discuss distinct properties of this Hamiltonian and its eigenstates. Specifically, in \cref{sec:wf_lowest}, we discuss the structure of lowest pair of well-localized wavefunctions for different Hamiltonian parameter configurations and distinguish them from those of an ordinary quadratic + quartic double-well potential. We present  our semiclassical analyses, namely a WKB analysis of the tunnel splitting in \cref{sec:WKB} and an overview of action quantization in \cref{sec:action} to discuss the construction of quantized orbits. We discuss in \cref{sec:spectrum} the robustness of the degeneracies in the squeeze-driven Kerr oscillator. \textcolor{black}{In \cref{model}, we present a simple Lindblad model to capture the qualitative features of the experimentally measured transverse relaxation lifetimes $T_X$ of the $\Delta$ variant of the Kerr-cat qubit}. 

Finally, in \cref{sec:tutorial} we present a self-contained tutorial and a concise introduction to the phase space formulation of quantum mechanics.

\section{Calibrating the squeezing drive amplitude $\epsilon_2$}
\label{sec:e2-calib}
\begin{figure*}
\includegraphics{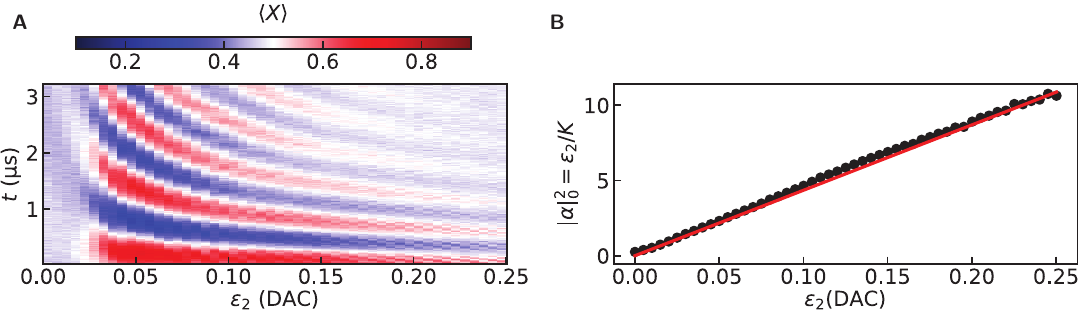}
\caption{\textbf{Calibrating $\epsilon_2$ with cat-Rabi oscillations.} \textbf{A} Color plot of $\langle X \rangle$ as a function of the the digital control amplitude (DAC) controlling the squeezing drive $\epsilon_2$ and duration of the Rabi drive. We find $\epsilon_x/2\pi = 144.93~\mathrm{kHz}$ using the relation between the Rabi amplitude and Rabi frequency for $\epsilon_2 = 0$, $\epsilon_x = \Omega_x (\epsilon_2 = 0)/2$. A plot of $|\alpha|_0^2 = \epsilon_2/K = \Omega_x^2/16 \epsilon_x^2$  \cite{grimm2020,frattini2022} as a function of $\epsilon_2$ in DAC units. A line fit gives us a calibration of $|\alpha|_0^2 = \epsilon_2/K$ as a function of the digital control amplitude (DAC) controlling the squeezing drive.}
\label{fig:e2-cal}
\end{figure*}

In this section, we present a measurement that provides an independent calibration of the squeezing drive amplitude $\epsilon_2$. The pulse sequence is the following: We turn on the squeezing drive at $\Delta = 0$, for a variable amount of time $t$ during which we also turn on a Rabi drive at amplitude $\epsilon_x$ and frequency $\omega_d/2 = \omega_a$. The squeezing drive stabilizes the Schr\"odinger cat states with well-defined parity, and the Rabi drive induces an oscillation in this cat-qubit. We perform this experiment for different values of $\epsilon_2$ and measure $\hat X = |\mathcal C^+\rangle\langle \mathcal C^-| + |\mathcal C^-\rangle\langle \mathcal C^+|$, where $|\mathcal C^\pm\rangle$ are the Schr\"odinger cat states. This protocol was introduced in \cite{grimm2020,frattini2022} and we refer the reader to these works for further details. The result of our experiment is shown in \Cref{fig:e2-cal}\textbf{A}. From this experimental data, we extract a Rabi oscillation frequency $\Omega_x$ that is related to the amplitude of the Rabi drive as $\epsilon_x = \Omega_x (\epsilon_2 = 0)/2$. The photon-number at $\Delta = 0$ $|\alpha|_0^2$ is related to $\epsilon_x$ and $\Omega_x$ as $|\alpha|^2_0 = \Omega_x^2/16 \epsilon_x^2$ \cite{grimm2020,frattini2022}. In \Cref{fig:e2-cal}\textbf{B}, we plot the experimental data and fit for the extracted photon-number as a function of the digital control amplitude (DAC). With this result, we have a calibration of $\epsilon_2$ as a function of the digital control amplitude (DAC) controlling the squeezing drive.
\section{Measuring the Kerr coefficient $K$}
\label{sec:K-calib}
\begin{figure}[h!]
\includegraphics{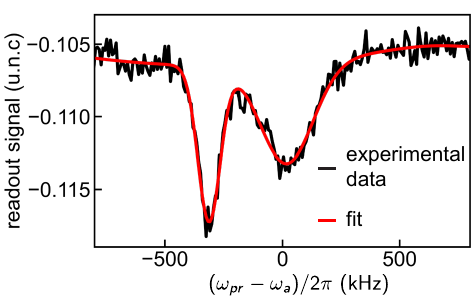}
\caption{Readout response as a function of the frequency of the saturation (probe) tone. The two readout signal dips in black correspond, from left to right, to the $gf/2$ transition, which is expected to occur at $(\omega_a - K)/2\pi$ and to the $ge$ transition of the SNAIL transmon, which is expected to occur at $(\omega_a)/2\pi$. Here, $gf/2$ refers to a transition induced by two photons from the probe. By fitting the experimental data, we find $K/2\pi = (329.73\pm4.30)~\mathrm{kHz}$.}
\label{fig:kerr-spec}
\end{figure}

In this section, we detail a measurement of the Kerr coefficient $K$ via saturation spectroscopy of the SNAIL transmon. This measurement is performed in the absence of the squeezing drive. In the following text, the letters $g, e,$ and $f$ index the ground, first excited, and second-excited states of the SNAIL transmon oscillator. In \Cref{fig:kerr-spec}, we plot the response of the readout as a function of a probe tone, whose frequency is $\omega_{pr}$, and which we vary around the $ge$ transition frequency of the SNAIL transmon oscillator $\omega_a$ corresponding to $\epsilon_2 = 0$. When the probe tone excites the oscillator, the readout signal due to the dispersive coupling \cite{blais2021} changes. The two dips in \Cref{fig:kerr-spec}, from left to right, correspond to a two-photon transition that excites the oscillator from $g$ to $f$ and to a resonant excitation of the oscillator from $g$ to $e$ respectively. The $gf/2$ and $ge$ resonances are located at  $(\omega_a - K)/2\pi$ and $\omega_a/2\pi$ respectively. Fitting the peaks and subtracting their locations yields a value of $K/2\pi = (329.73\pm4.30)~\mathrm{kHz}$. This value is consistent with the value of $K/2\pi = 316.83~\mathrm{kHz},$ where the latter is extracted from Figure 1\textbf{E} in the main text and is the value for $K$ used throughout the article.

\pagebreak

\section{Exponential reduction of tunneling with $\epsilon_2$}
\label{sec:exp-red-data}

\begin{figure*}
\includegraphics{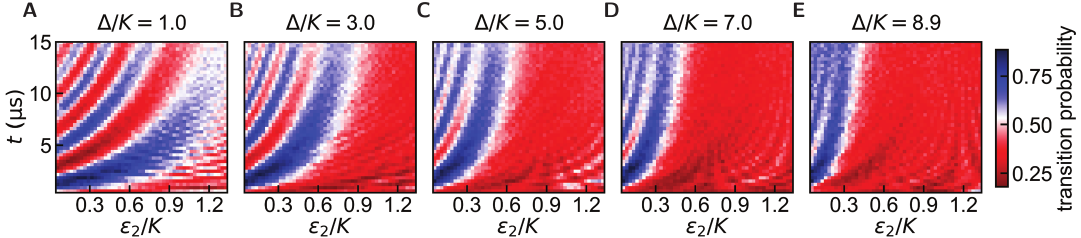}
\caption{\textbf{Tunnel-driven Rabi oscillations in the ground state manifold and its exponential reduction as a function of $\epsilon_2$; raw data.} The transition probability as a function of $\epsilon_2/K$ and time $t$ for \textbf{A} $\Delta/K = 1$, \textbf{B} $\Delta/K = 3$, \textbf{C} $\Delta/K = 5$, \textbf{D} $\Delta/K = 7$, and \textbf{E} $\Delta/K = 9$ respectively. This corresponds to the condition of constructive interference of tunneling to occur. By progressively increasing $\epsilon_2$, there is a clear overall continuous reduction of the tunnel-driven Rabi oscillations.}
\label{fig:exp-supp-e2}
\end{figure*}

In the main text, we claim that tunneling in the ground state manifold is overall continuously reduced with $\epsilon_2$. The parameter $\epsilon_2$ controls the barrier height, which is given as $(\Delta + 2\epsilon_2)^2/4K$ in the double-node phase and $2 \Delta \epsilon_2/K$ in the triple-node phase. Moreover, continuing this reasoning, the larger the detuning, the faster the tunneling reduction as a function of $\epsilon_2$. In \Cref{fig:exp-supp-e2}, we present raw data to further support this claim. We present the measurement protocol in Figure 1\textbf{C} of the main text and recall it for the sake of completeness. First, we prepare, by measurement, a steady-state localized in one of the wells. Following this, we adiabatically lower the squeezing drive amplitude $\epsilon_2$. Lowering the value of $\epsilon_2$ reduces the barrier depth, and thus the tunnel effect becomes observable.  We then wait for a variable amount of time before adiabatically re-raising $\epsilon_2$ to its initial value and finally do which-well readout.

In \Cref{fig:exp-supp-e2}, we present the measured transition probability as a function of $\epsilon_2$ for \textbf{A} $\Delta/K = 1$, \textbf{B} $\Delta/K = 3$, \textbf{C} $\Delta/K = 5$, \textbf{D} $\Delta/K = 7$, and \textbf{E} $\Delta/K = 9$ respectively. It is clear from the data that the Rabi-frequency is overall continuously reduced with $\epsilon_2$ and moreover, increasing $\Delta/K$ reduces the Rabi frequency further. We plot in Figure 2 of the main text the extracted tunneling amplitude $|\delta E|$ from our data by fitting the oscillation frequency with an exponentially decaying sinusoid. We find that the extracted tunneling amplitude is in excellent agreement with an exact diagonalization of the static effective Hamiltonian and in good agreement with a WKB prediction of the tunnel splitting within the expected regime of validity. See \Cref{fig:WKB-diag} for more details.

\section{Transverse relaxation lifetime $T_X$ measurements}
\label{sec:lifetime}

\begin{figure*}
\includegraphics{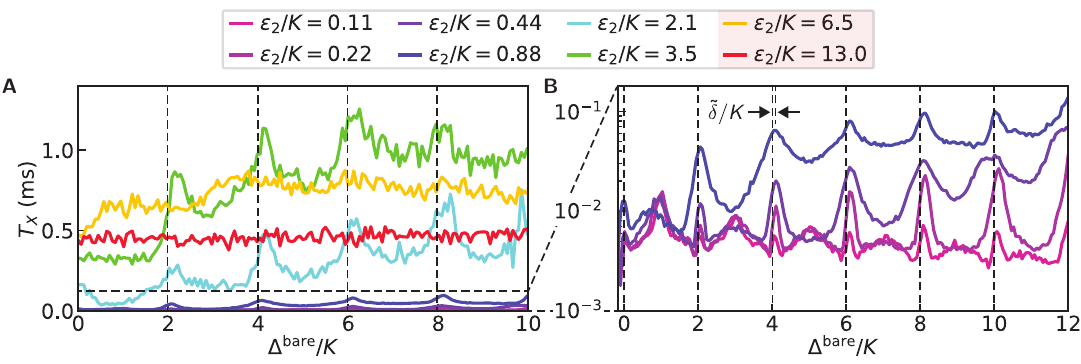}
\caption{\textbf{Measurement of $T_X$ as a function of $\Delta^{\mathrm{bare}} = \omega_a - \omega_d/2$ for representative values of squeezing drive amplitude $\epsilon_2$}. The measurement protocol is shown in Figure 3\textbf{E} of the main text. We observe a degradation of $T_X$ with increasing $\epsilon_2$, as indicated by red boxes in the legend, and we show representative measurements here. On the other hand, we see no degradation of $T_X$ with increasing $\Delta$. This measurement indicates that $\Delta$ might be a more effective knob to increase $T_X$ than $\epsilon_2$ for cat-states with large photon-number \cite{roberts2020}. \textbf{B} Low-power lifetime extracted from time-resolved measurements. The data is presented as Figure 1\textbf{F} in the main text.}
\label{fig:lifetimes-all}
\end{figure*}
 
In \Cref{fig:lifetimes-all}, we plot the transverse relaxation lifetime $T_X$ as a function of $\Delta^{\mathrm{bare}} = \omega_a - \omega_d/2$ for different values of $\epsilon_2$. Note that the photon-number at $\Delta = 0$ is given by $|\alpha|_0^2 = \epsilon_2/K$. Importantly, for large photon-numbers $\epsilon_2/K \gtrsim 6.5$, we see that the peaks in lifetime start plateauing and even dropping. This effect is not captured by an ordinary model of the Lindblad master equation as we discuss in \cref{model}. The degradation of the $T_1$ with readout power has been observed for transmon qubits \cite{sank2016}. But other drive-induced effects such as multiphoton nonlinear resonances are present in transmons and disentangling these various sources of lifetime degradation is nontrivial \cite{sank2016,blais2021,Shillito2022,Cohen2022,petrescu2020}. These spurious nonlinear resonances are largely absent in this our SNAIL conducting circuit for values of $\epsilon_2/K \lesssim 5$, thanks to negligible Kerr and stark shifts, but may plague our system for larger mean-photon numbers. Due to this reasoning, $\Delta$ might be a more effective knob to create states with large photon number \cite{roberts2020}. Finally, the squeeze-driven Kerr oscillator provides a perfect platform to investigate lifetime degradation under drives.

\pagebreak
\section{Notation}
\label{sec:notation}In this work, we note $\hat X$ and $\hat P$ the \textit{position}-like and \textit{momentum}-like coordinates with $[\hat X,\hat P]=i\hbar$. We build the dimensionless quadratures by introducing the zero point spread of the coordinates as $X_{\textrm{zps}}$ and $P_{\textrm{zps}}$, respecting $X_{\textrm{zps}}P_{\textrm{zps}} = \hbar/2$. We further introduce the complex notation for the dimensionless quadratures as $\hat a =  (\hat X/X_{\textrm{zps}} + i \hat P / P_{\textrm{zps}})/2$ and its conjugate operator $\hat{a}^\dagger$, where $[\hat{a}, \hat{a}^\dagger] = 1$ and introduce the rescaled phase space quadratures as $\hat{x} = \sqrt{\lambda/2} \hat X/X_{\textrm{zps}} = \sqrt{\lambda/2} (\hat{a} + \hat{a}^\dagger)$ and $\hat{p} = \sqrt{\lambda/2}\hat P/P_{\textrm{zps}} = -i\sqrt{\lambda/2} (\hat{a} - \hat{a}^\dagger)$, where $[\hat{x}, \hat{p}] = i \lambda$. These choices induce the definitions $x_{\textrm{zps}}=p_{\textrm{zps}}=\sqrt{\lambda/2}$. Conversely, we have $\hat{a} = (\hat{x} + i \hat{p})/\sqrt{2\lambda}$. At this point, $\lambda$ is a dimensionless rescaling parameter. We will connect it with the Hamiltonian parameters later, while discussing the classical limit ($\lambda\rightarrow 0$) of our system, and thereby give it physical significance. It is also useful to compare our results with those of \cite{marthaler2007}, who have performed a WKB analysis of a driven oscillator. Thus, unless otherwise specified, $\lambda$ should be taken equal to unity $\lambda =1$.

For a mechanical oscillator with mass $m$ and spring-constant $k$, the small-oscillation frequency is $\omega_o = \sqrt{k/m}$ and the impedance is $Z_o = 1/\sqrt{km}$. With this, we have $X_{\mathrm{zps}} = \sqrt{\hbar Z_o/2}$ and $P_{\mathrm{zps}} = \sqrt{\hbar/2 Z_o}$.
We further remark that there is a direct correspondence between the mechanical harmonic oscillator and a linear LC circuit oscillator \cite{devoret1995,girvin2014,blais2021} under the following relations. The mechanical position coordinate $\hat X$ corresponds to the circuit flux $\hat \Phi$, the mechanical momentum $\hat P$ corresponds to the circuit charge $\hat Q$, where $[\hat \Phi,\hat Q]=i\hbar$, the mechanical oscillator frequency $\omega_o = \sqrt{k/m}$ corresponds to the circuit oscillator frequency $\omega_o = 1/\sqrt{LC}$ and the mechanical oscillator impedance $Z_o=1/\sqrt{km}$ corresponds to the circuit oscillator impedance $Z_o = \sqrt{L /C}$ which amounts to the identification of the mechanical mass $m$ with the circuit capacitance $C$ and the spring constant $k$ with the inverse inductance $1/L$. The expressions for the zero point spreads are given by $\Phi_{\mathrm{zps}} = \sqrt{\hbar Z_o/2}$ and $Q_{\mathrm{zps}} = \sqrt{\hbar /2  Z_o}$. In circuits, it is customary to introduce \cite{devoret2021,blais2021} the reduced flux and charge coordinates: $\hat \varphi = \sqrt{\lambda}2 \pi  \hat \Phi/\Phi_0$ and $\hat N = \sqrt{\lambda}\hat Q/2e$  so that $[\hat \varphi, \hat N] = i\lambda$, where $e$ is the charge quantum, and $\Phi_0 = h/2e$ is the magnetic flux quantum.\footnote{Note that in this case the non-dimensionalization of variables is done by fundamental constants and not by linear properties of the oscillator. This comes at the price of a slight notation asymmetry over the reduced operators the electric and mechanical oscillators.} Their respective zero point spreads $\varphi_{\mathrm{zps}} =  \sqrt{\lambda}2 \pi \Phi_{\mathrm{zps}}/\Phi_0$ and $N_{\mathrm{zps}}=  \sqrt{\lambda}Q_{\mathrm{zps}}/2e$,  and are related to the rescaled complex coordinate operators by $\hat \varphi = \varphi_{\mathrm{zps}} (\hat a^\dagger + \hat a)$ and $\hat N = -i N_{\mathrm{zps}} (\hat a-\hat a^\dagger)$ and $\varphi_{\mathrm{zps}} N_{\mathrm{zps}} = \lambda/2$. We summarize this notation in the following table

\begin{table}
\begin{tabular}{c|c}
Mechanical oscillator & Circuit oscillator  \\
\hline\hline
$\hat{X};\; \hat P$ & $\hat{\Phi};\;\hat Q$ \\
\hline$[\hat{X}, \hat{P}]=i \hbar$ & $[\hat{\Phi}, \hat{Q}]=i \hbar$ \\
\hline$\omega_o= \sqrt{k/m}$ & $ \omega_o = 1/\sqrt{LC}$ \\
\hline$Z_o=1 / \sqrt{k m}$ & $Z_o=\sqrt{L/C}$ \\
\hline$X_{\textrm{zps}}=\sqrt{\hbar Z_o/2}$; & $\Phi_{\textrm{zps}}=\sqrt{\hbar Z_o/2};$ \\
$P_{\textrm{zps}}=\sqrt{\hbar/2 Z_o}$ & $Q_{\textrm{zps}}=\sqrt{\hbar/2 Z_o}$ \\
$\Rightarrow X_{\textrm{zps}} P_{\textrm{zps}}=\hbar/2$ & $\Rightarrow \Phi_{\textrm{zps}} Q_{\textrm{zps}}=\hbar/2$ \\
\hline $\hat{a}=\frac{1}{2}\left(\frac{\hat{X}}{X_{\textrm{zps}}}+i \frac{\hat{P}}{P_{\textrm{zps}}}\right)$ & $\hat{a}=\frac{1}{2}\left(\frac{\hat{\Phi}}{\Phi_{\textrm{zps}}}+i \frac{\hat{Q}}{Q_{\textrm{zps}}}\right)$ \\
\hline$\hat{X} = X_{\textrm{zps}}\left(\hat{a}+\hat{a}^{\dagger}\right)$ & $\hat{\Phi}=\Phi_{\textrm{zps}}\left(\hat{a}+\hat{a}^{\dagger}\right)$ \\
\hline$\hat{P}=-i P_{\textrm{zps}}\left(\hat{a}-\hat{a}^{\dagger}\right)$ & $\hat{Q}=-i Q_{\textrm{zps}} \left(\hat{a}-\hat{a}^{\dagger}\right)$\\
\hline $\left[\hat{a}, \hat{a}^{\dagger}\right]=1$ & $\left[\hat{a}, \hat{a}^{\dagger}\right]=1$\\
\hline
\hline
$\hat{x}=\sqrt{\frac{\lambda}{2}} \frac{\hat{X}}{X_{\textrm{zps}}}=x_{\textrm{zps}}\left(\hat{a}+\hat{a}^{\dagger}\right)$ & $\hat{\varphi}=\sqrt{\lambda} 2 \pi \frac{\hat{\Phi}}{\Phi_0}=\varphi_{\textrm{zps}}\left(\hat{a}+\hat{a}^{\dagger}\right)$ \\
\hline$\hat{p}=\sqrt{\frac{\lambda}{2}} \frac{\hat{P}}{P_{\textrm{zps}}}=-i p_{\textrm{zps}}\left(\hat{a}-\hat{a}^{\dagger}\right)$ & $\hat{N}=\sqrt{\lambda} \frac{\hat{Q}}{2 e}=-i N_{\textrm{zps}} \left(\hat{a}-\hat{a}^{\dagger}\right)$\\
\hline$x_{\textrm{zps}}=p_{\textrm{zps}}=\sqrt{\lambda/2}$ & $\varphi_{\textrm{zps}}=2 \pi \sqrt{\lambda} \frac{\Phi_{\textrm{zps}}}{\Phi_0} ;  N_{\textrm{zps}}=\sqrt{\lambda} \frac{Q_{\textrm{zps}}}{2 e}$ \\
$\Rightarrow x_{\textrm{zps}} p_{\textrm{zps}}=\lambda/2$ &$\Rightarrow \varphi_{\textrm{zps}} N_{\textrm{zps}}=\lambda/2$ \\
\hline$[\hat{x}, \hat{p}]=i \lambda$ & $[\hat{\varphi}, \hat{N}]=i \lambda$ \\
\hline $\hat{a}=\frac{1}{2}\left(\frac{\hat{x}}{x_{\textrm{zps}}}+i \frac{\hat{p}}{p_{\textrm{zps}}}\right)$ & $\hat{a}=\frac{1}{2}\left(\frac{\hat{\varphi}}{\varphi_{\textrm{zps}}}+i \frac{\hat{N}}{N_{\textrm{zps}}}\right)$ \\
\hline
$\hat a =(\hat{x}+i \hat{p})/\sqrt{2 \lambda}$ & $\hat{a}=\left(\sqrt{ \frac{\lambda}{2}}\frac{\hat{\varphi}}{\varphi_{\textrm{zps}}}+i \sqrt{ \frac{\lambda}{2}}\frac{\hat{N}}{N_{\textrm{zps}}}\right)/\sqrt{2 \lambda}$\\
\end{tabular}
\end{table}

\section{The relationship between different squeeze-driven Kerr oscillator models in the literature and their classical limit}
\label{sec:relation}

In 1993, Wielinga and Milburn \cite{Wielinga1993} proposed a quantum optical model that they called \textit{the dynamical equivalent of the double-well potential}. The interest of the problem, to them, was that their model exhibited a double-well structure in phase space, and quantum mechanical ground state tunneling between them. The Hamiltonian they addressed is
\begin{equation}
    \hat H_{\textrm{WM}} = -K (\hat a ^{\dagger}\hat a)^2 + \epsilon_2(\hat a ^{\dagger 2}+\hat a^2).
\end{equation}

In 2017, the theoretical discovery of the Kerr-cat qubit by Puri, Boutin, and Blais \cite{puri2017} relied on the fact that the ground states of 

\begin{equation}
\label{eq:SP}
    \hat H_{\textrm{PBB}}  = -K \hat a ^{\dagger 2}\hat a^2 + \epsilon_2(\hat a ^{\dagger 2}+\hat a^2)
\end{equation}
are fundamentally degenerate and exhibit no tunneling between two wells found in the classical limit (see also \cite{cochrane1999}). This property can be understood by writing \cref{eq:SP} into the factorized form \cite{puri2017}

\begin{equation}
\label{eq:SPfactorization}
    \hat H_{\textrm{PBB}}  = -K (\hat a ^{\dagger 2}-\epsilon_2/K)(\hat a ^{ 2}-\epsilon_2/K),
\end{equation}
from which it is evident that the two coherent states $|\pm \alpha \rangle$ with $\alpha = \sqrt{\epsilon_2/K}$, which are the eigenstates of the annihilation operator $\hat{a}$, are also degenerate eigenstates of \cref{eq:SPfactorization}.  Since \cref{eq:SPfactorization} is negative-semidefinite and $\hat H_{\textrm{PBB}}|\pm \alpha \rangle = 0$, these states are the ground states.

Note that the Hamiltonians $\hat H_{\textrm{WM}}$ and $\hat H_{\textrm{PBB}}$ differ only by a commutator.  Their shared classical limit can be written as

\begin{align}
\label{eq:H-cl-4}
\begin{split}
     H_{\textrm{cl}} &= -K  a ^{* 2} a^2 + \epsilon_2( a ^{* 2}+ a^2)\\
    &=-K \left(\frac{x^2+p^2}{2}\right)^2 + \epsilon_2 (x^2-p^2).
\end{split}
\end{align}

\begin{figure*}
\includegraphics{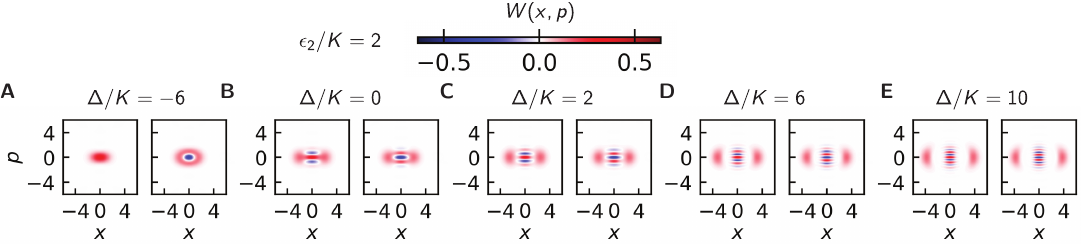}
\caption{\textbf{Lowest eigen-manifold of the squeeze-driven Kerr oscillator}  Wigner functions of lowest pair of eigenstates of  \cref{eq:H-eff} (top row) for $\epsilon_2/K = 2$ and  \textbf{A} $\Delta/K =-6$, \textbf{B} $\Delta/K =0$, \textbf{C} $\Delta/K =2$, and \textbf{D} $\Delta/K =6$ and $\Delta/K = 10$ respectively. $\Delta/K \ll 0$, the eigenstates are squeezed. For $\Delta/K \ll 0$, increasing $\Delta/K$ yields Schr\"{o}dinger cat states with increasing photon number. This phenomenon manifests as the monotonically growing baseline in transverse relaxation lifetime $T_X$ in Figure 3 of the main text.}
\label{fig:wig-l}
\end{figure*}

\begin{figure*}
\includegraphics{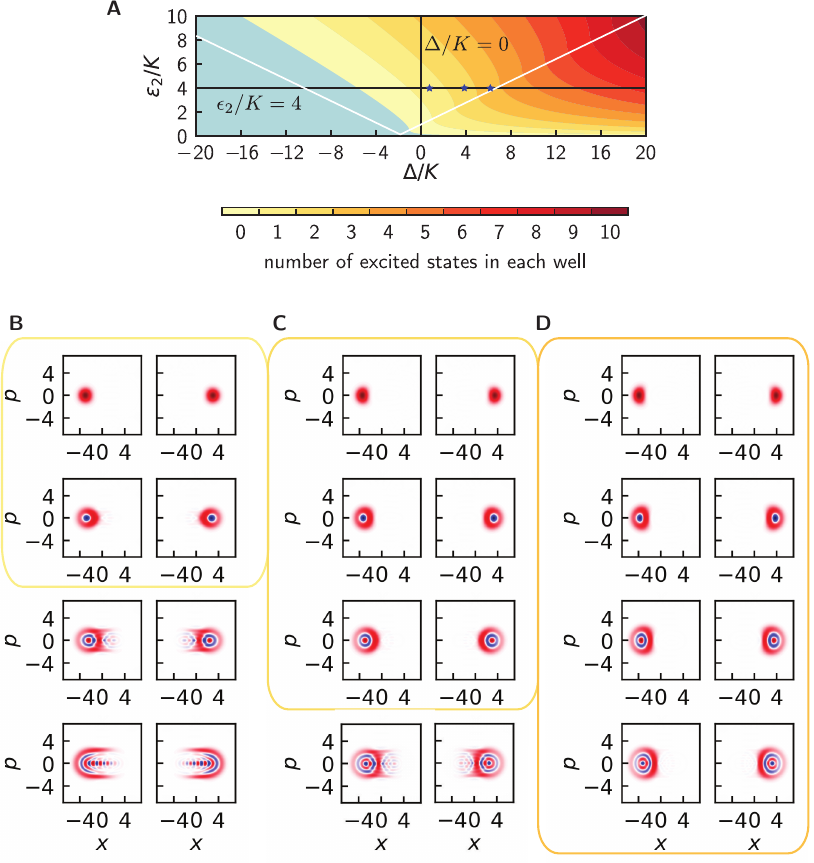}
\caption{\textbf{Excited well states in the squeeze-driven Kerr oscillator} Localized ground and excited states in the squeeze-driven Kerr oscillator. \textbf{A} Period doubling phase diagram with equi-state contours. \textbf{B} - \textbf{E}. Wigner functions of exact eigenstates' superpositions, corresponding to localized states, for $\epsilon_2/K = 4$, and \textbf{B} $\Delta/K = 1$, \textbf{C} $\Delta/K = 4$, \textbf{D} $\Delta/K = 7$. The action quantization formulation, detailed in \cref{sec:semi} and summarized by \cref{eq:Nstates}, predicts $\textbf{B}\, 1, \textbf{C}\, 2, \textbf{D}\, 3$ excited states in each well. The Wigner functions of states outside this window are seen to have support in the other well too, and larger $\Delta$ helps localize them, thus validating the semiclassical picture discussed in \cref{sec:semi} quantum mechanically.}
\label{fig:wig-h}
\end{figure*}

Since $\hat{a}^{\dagger 2} \hat{a}^2 - (\hat{a}^{\dagger } \hat{a})^2 = \hat{a}^\dagger \hat{a}$, we cast the Hamiltonian
\begin{equation}
\label{eq:H-eff}
    \hat H = \Delta\hat a ^{\dagger }\hat a -K \hat a ^{\dagger 2}\hat a^2 + \epsilon_2(\hat a ^{\dagger 2}+\hat a^2),
\end{equation}
where we identify that $\hat H_{\textrm{PBB}}$ and $\hat H_{\textrm{WM}}$ are specific instances of \cref{eq:H-eff} with $\hat H_{\textrm{PBB}} = \hat{H}|_{\Delta = 0}$ and $\hat H_{\textrm{WM}} = \hat{H}|_{\Delta = -K}$. Note that taking $\Delta\neq 0$ breaks the simple factorization condition of \cref{eq:SPfactorization}. Indeed, the presence of the $\hat{a}^\dagger \hat{a}$ term is the cause of ground state tunneling in $\hat H_{\textrm{WM}}$, and its absence is the cause of the complete coherent cancellation of tunneling in $\hat H_{\textrm{PBB}}$. The lowest eigen-manifold of \cref{eq:H-eff} is plotted in \Cref{fig:wig-l} while the  excited state manifold of \cref{eq:H-eff} is plotted in \Cref{fig:wig-h}.


In 2007, Marthaler and Dykman \cite{marthaler2007} treated a Hamiltonian similar to \cref{eq:H-eff}, where $\Delta$ was kept free for a fixed $\epsilon_2$. This led to their prediction of periodic cancellation of tunneling amplitude for the ground state manifold as a function of $\Delta$. Their work inspired our experiment shown in Figure 1 of the main text.  We discuss in detail the mapping of their problem to ours in \cref{sec:semi}. 

In the following text, we discuss the quantum phase space representation of  \cref{eq:H-eff}.

\section{Phase space formulations of our effective Hamiltonian}
\label{sec:rep-H-H}
Let us reconsider \cref{eq:H-eff}. For its derivation starting from the circuit Hamiltonian, see appendix A of \cite{frattini2022}.

We obtain the phase space formulation of \cref{eq:H-eff} by taking the invertible Wigner transform \cite{curtright2013} $\mathfrak W$ as
\begin{align}
\label{eq:Wig-KC}
\begin{split}
    &\hat{x} \rightarrow \mathfrak W\{\hat x \} = x;\qquad 
    \hat{p} \rightarrow \mathfrak W \{ \hat p \}= p; \qquad \\
    & \hat{a} \rightarrow \mathfrak W \{ \hat{a} \} = a = (x + i p)/\sqrt{2 \lambda};  \quad \hat{a}^\dagger \rightarrow \mathfrak W  \{ \hat{a}^\dagger \} = a^*; \\
    &\hat{a}^\dagger \hat{a} \rightarrow a^* \star a  = a^*a - \frac{1}{2} = \frac{x^2 + p^2}{2 \lambda}- \frac{1}{2}; \\
    & \hat{a}^{\dagger 2} \hat{a}^2 \rightarrow a^{*2} \star a^2 = a^{*2} a^2 - 2a^* a + \frac{1}{2} \\  & \qquad \qquad \qquad \quad = \frac{(x^2 + p^2)^2}{4 \lambda^2} - \frac{ (x^2 + p^2)}{\lambda}+ \frac{1}{2}; \\
    &\hat{a}^{\dagger 2} +  \hat{a}^2 \rightarrow a^{* 2} +  a^2  = \frac{(x^2 - p^2)}{\lambda},
\end{split}
\end{align}
where the Groenewold star product \cite{groenewold1946} is given by $\mathfrak W \{ \hat A\hat B \} = A \star B = A \exp\left(\frac{1}{2}(\overleftarrow{\partial}_{a} \overrightarrow \partial_{ a^*}-\overleftarrow{\partial}_{ a^*} \overrightarrow \partial_{ a})\right) B$. The Moyal bracket \cite{moyal1949,curtright2013} over $a$ and $a^*$ is defined as $ \{\!\!\{ A, B\}\!\!\}_{a, a^*} =  A \star B - B \star A$ so that we have $\{\!\!\{  a,  a^*\}\!\!\} = 1$.
For a pedagogical exposition on the phase space formulation of quantum mechanics, we refer the reader to \cref{sec:tutorial} and \cite{hillery1984,case2008,curtright2013}. With \cref{eq:Wig-KC}, we write \cref{eq:H-eff} in the phase space formulation of quantum mechanics, up-to coordinate-independent terms, as
\begin{align}
\begin{split}
\label{eq:H-eff-ps}
H = (\Delta + 2K) \left(\frac{x^2 + p^2}{2 \lambda } \right) - K \left(\frac{x^2 + p^2}{2  \lambda }\right)^2  + \epsilon_2\left(\frac{x^2 - p^2}{ \lambda }\right).
\end{split}
\end{align}
Note that \cref{eq:H-eff-ps} is not equal to \cref{eq:H-cl-4} even when $\Delta = 0$ and $\lambda = 1$.
We further rescale \cref{eq:H-eff-ps} by $-K /\lambda^2$ so as to have a coefficient of order 1 for the nonlinear term and rearrange \cref{eq:H-eff-ps} as

\begin{align}
\begin{split}
\label{eq:H-eff-ps-2}
\frac{-H \lambda^2}{K} = \left(\frac{x^2 + p^2}{2}\right)^2 - \frac{2 \epsilon_2 \lambda}{K } \frac{x^2}{2} \left( 1 + \frac{(\Delta + 2 K)}{2 \epsilon_2}\right)  + \frac{2 \epsilon_2  \lambda}{K} \frac{p^2}{2} \left( 1 - \frac{(\Delta + 2 K )}{2 \epsilon_2}\right).
\end{split}
\end{align}

By choosing the scale of phase space $\lambda = K/2\epsilon_2$ \cref{eq:H-eff-ps-2} becomes
\begin{align}
\begin{split}
\label{eq:H-eff-ps-3}
\frac{-H \lambda^2}{K } = \left(\frac{x^2 + p^2}{2}\right)^2  -  \frac{x^2}{2} \left( 1 + \frac{\Delta}{2 \epsilon_2} + 2 \lambda  \right)  + \frac{p^2}{2} \left( 1 - \frac{\Delta}{2 \epsilon_2} - 2 \lambda \right).
\end{split}
\end{align}

The term proportional to $\lambda$ in \cref{eq:H-eff-ps-3} involves a commutator, and corresponds to the Lamb shift. The classical limit then consist in dropping this term. This is valid for $\lambda \ll \mathrm{min}(\Delta/2\epsilon_2,\;1)$. This translates to $\Delta/K,\; \epsilon_2/K \gg 1$. In this limit, the WKB approximation is valid to treat Eq. (1) in the main text.

The interpretation of the Hamiltonian classical limit is that the elementary action element $\lambda$ in the phase space defined by $x$ and $p$ must be much smaller than the typical dimensionless action of the system determined by the well-size parameters: $\Delta/K$ and $\epsilon_2/K$. As we discuss in what follows (see section \cref{sec:meta-cl-props} and \cite{frattini2022}), under the condition $\Delta/K,\; \epsilon_2/K \gg 1$, the wells of the Hamiltonian are large in the sense that they encompass many action quanta $\lambda$. Finally, note that for $\lambda\approx 1$ the classical treatment should not hold.

We call the surface for $H$ in \cref{eq:H-eff-ps-3} the metapotential of the squeeze-driven Kerr oscillator, and the classical limit for $H$ in \cref{eq:H-eff-ps-cl} as the classical metapotential surface. Furthermore, as customary, we plot $-H$ rather than $H$ to respect the familiar notion that in the presence of dissipation, stable equilibria correspond to well-bottoms rather than hill-tops.

\begin{widetext}

\subsubsection{Properties of the metapotential surface}
\label{sec:meta-cl-props}
In the table below, we examine the properties of the metapotential surface. For details on the number of levels inside the well, which we obtain via action quantization following the prescription of Einstein-Brillouin-Keller (EBK) \cite{gutzwiller2013}, see \cref{sec:action}.
\begin{center}
\begin{tabular}{|c| c| c| c|}
\hline
    Phase $\rightarrow$ & Double-node $-2 \epsilon_2 \le \Delta + 2K \le 2 \epsilon_2$ & Triple-node $\Delta + 2K > 2\epsilon_2$ \\
    $\downarrow$ Parameter & & \\
    ($x, p$ phase space) & & \\
    \hline
    Area  & $\frac{(\Delta + 2K)}{K} \arccos{\left(\frac{-(\Delta + 2K)}{2\epsilon_2}\right)} + \frac{2\epsilon_2}{K} \sqrt{1 - \left( \frac{(\Delta + 2K)}{2 \epsilon_2} \right)^2}$ & $\frac{4\epsilon_2}{K} \sqrt{\frac{(\Delta + 2K)}{2\epsilon_2} - 1} + \frac{2 (\Delta + 2K)}{K}\arcsin{\left(\sqrt{\frac{2\epsilon_2}{(\Delta + 2K)}} \right)}$ \\
    \hline
    Levels per well (\#)  & area$/2\pi-1/2$ & area$/2\pi-1/2$ \\
    \hline
    Approximation of \# &  $\frac{(\Delta + 2K)/K}{2} + \frac{\epsilon_2/K}{\pi} - \frac{1}{2}$ &  $\frac{\sqrt{8 \epsilon_2 (\Delta + 2K)}}{K \pi} - \frac{1}{2}$\\
    \hline
    Distance b/w nodes & $2\sqrt{\frac{(\Delta + 2K) + 2 \epsilon_2}{K}}$ & $2\sqrt{\frac{(\Delta + 2K) + 2 \epsilon_2}{K}}$ \\
    \hline
    Distance b/w saddles  & $0$ & $2\sqrt{\frac{(\Delta + 2K) - 2 \epsilon_2}{K}}$ \\
    \hline
    Depth of nodes & $\frac{(\Delta + 2K + 2 \epsilon_2)^2}{4 K}$ & $\frac{(\Delta + 2K + 2 \epsilon_2)^2}{4 K}$ \\
    \hline
    Depth of saddles & 0 & $\frac{(\Delta + 2K - 2 \epsilon_2)^2}{4 K}$ \\
    \hline
    Depth of barrier & $\frac{(\Delta + 2K + 2\epsilon_2)^2}{4K}$ & $\frac{2 (\Delta + 2K) \epsilon_2}{K}$\\
    \hline
\end{tabular}
\end{center}
\end{widetext}

\subsection{Wavefunctions of localized well states}
\label{sec:wf_lowest}
In this section, we examine closely the wave functions of the squeeze-driven Kerr oscillator in the classically forbidden region and contrast them with those of an ordinary quadratic + quartic potential. We define the ordinary double-well Hamiltonian as
\begin{align*}
    H = \frac{p^2}{2} + V(x), \qquad \mathrm{with} \qquad V(x) = -\frac{k_2}{2} x^2 + \frac{k_4}{4} x^4,
\end{align*}
where $k_2, \;k_4 > 0$. This potential has a saddle at $x_s = 0$, with $V(x_s) = 0$ and nodes at $x_n = \pm \sqrt{k_2/k_4}$ with the left and right well depth given by $V(x_n) = -k_2^2/(4 k_4)$. The barrier height is given by $V(x_n) - V(x_s) = k_2^2/(4k_4)$.

The study of tunneling usually begins by considering a localized wave packet in one well, which is written as the superposition of the wavefunctions of the two lowest laying energy states $\psi_+$ and $\psi_-$.\footnote{From a perturbation theory point of view this corresponds to the bonding and anti-bonding of the decoupled well states \cite{landau1991}. The zero point energy of the individual wells, in the absence of tunneling, is $ E_0 = \sqrt{k_2}/2$. In the presence of tunneling the system's energies can be approximated by $E^{\pm}=E_0\pm \delta E$. } Their energy difference is denoted by $\delta E = E^+-E^-$ and the left- and right-localized wavefunctions read
\begin{align}
    \psi_l = \frac{\psi_+ + \psi_-}{\sqrt{2}} \qquad \psi_r = \frac{\psi_+ - \psi_-}{\sqrt{2}}.
\end{align}

\begin{figure*}
\includegraphics{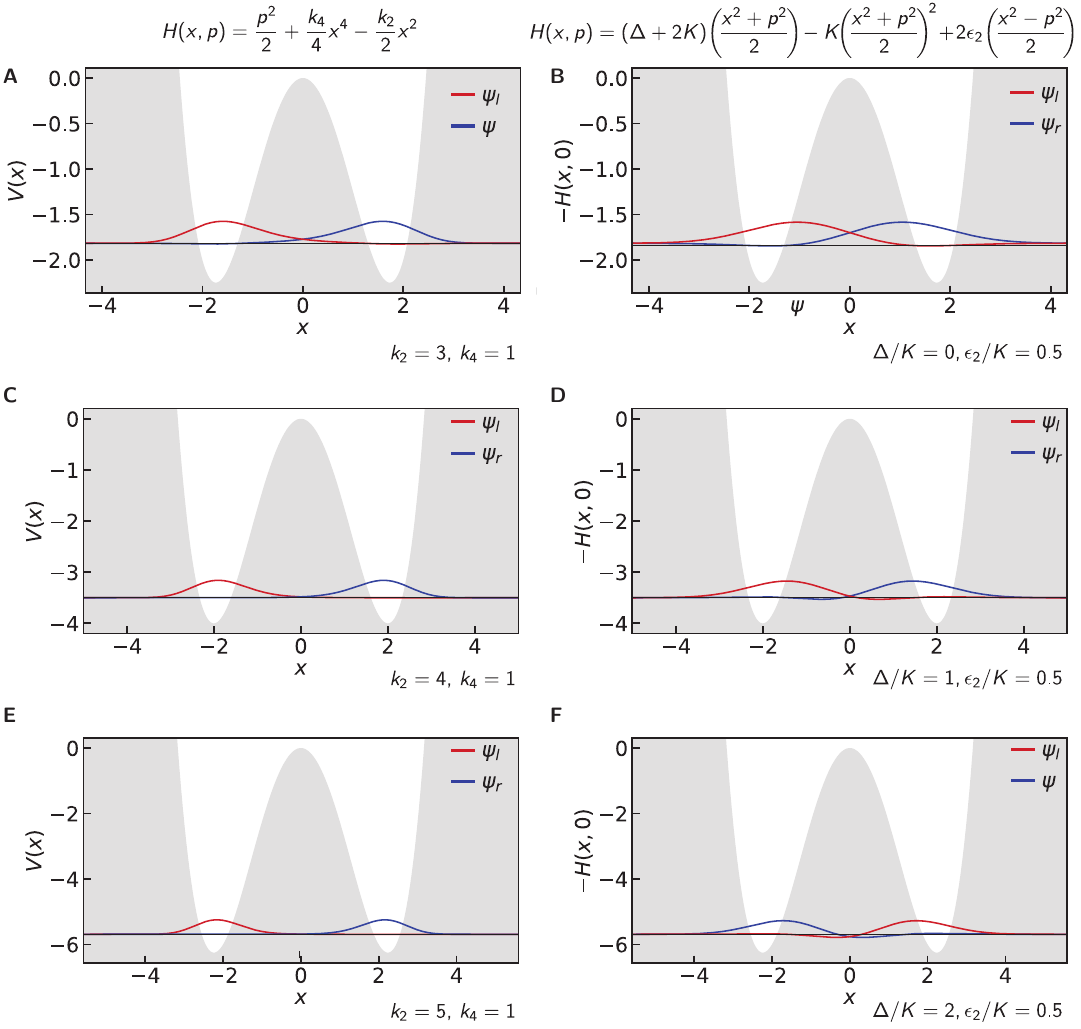}
\caption{\textbf{Localized wavefunctions of the ground state manifold in the position basis in $\textbf{A}, \textbf{C}, \textbf{E}$ an ordinary double well potential and in \textbf{B}, \textbf{D}, \textbf{F} for a squeeze-driven Kerr oscillator.} The Hamiltonian parameters in \textbf{A}, \textbf{C}, and \textbf{E} have been chosen to produce a double-well with the same depth and the well separation as those of \textbf{B}, \textbf{D}, and \textbf{F} respectively. The value of $\Delta/K$ is chosen to be \textbf{B} $\Delta/K = 0$,  \textbf{D} $\Delta/K = 1$, and  \textbf{F} $\Delta/K = 2$ corresponding to the destructive, constructive, and destructive interference of tunneling respectively. In the right panel, oscillations accompany decay of the wavefunction in the classically forbidden region, marked in grey. In the left panel, the wavefunction exhibits pure decay in the classically forbidden region. In this sense, the cancellation of tunneling amplitude in fig 1 of the main text can be understood as the destructive interference of the wavefunction in the classically forbidden region of the squeeze-driven Kerr oscillator. In \cite{marthaler2007}, Marthaler and Dykman found an analytical expression for the WKB tunnel splitting of the ground state manifold. See \cref{sec:WKB} for the WKB expressions for the tunnel splitting and \Cref{fig:WKB-diag}\textbf{B} and \textbf{D} for comparisons of the extracted tunnel splitting from experiment with their WKB theory.}
\label{fig:wf_lowest}
\end{figure*}

On the left column of \Cref{fig:wf_lowest}, we plot the left and right-localized wavefunctions in red and blue respectively for \textbf{A} $k_2 = 3, \, k_4 = 1$, \textbf{B} $k_2 = 2, \, k_4 = 1$, and \textbf{B} $k_2 = 4, \, k_4 = 2$ respectively. The wavefunctions are computed by numerical diagonalization of the Hamiltonian. In the classically forbidden region, as one should expect, the wavefunctions display evanescent decay \cite{griffiths2018}.

In the right column of \Cref{fig:wf_lowest}, we contrast the localized wavefunctions of the ordinary double-well potential with those of the squeeze-driven Kerr oscillator. The parameters $\Delta$, $K$, and $\epsilon_2$ were chosen so that a cut of the effective Hamiltonian surface at $p=0$ yields an identical double-well potential as the left column. The wavefunctions of the full squeeze-driven Kerr oscillator are computed numerically. Importantly, \textbf{B}, \textbf{D}, and \textbf{F} show the localized wavefunctions for $\Delta/K = 0, 1,$ and $2$ respectively, corresponding to the destructive, constructive, and again destructive interference. Interestingly, in the classically forbidden region, in \textbf{B} and \textbf{D}, oscillations accompany decay in the wavefunction \cite{marthaler2006,marthaler2007}. This is due to the underlying driven nature of our system, providing a quartic term in momentum, which here reflects in the oscillatory nature of the wavefucntions in the classically forbidden region.

\section{Semiclassical analysis}
\label{sec:semi}
In \cite{marthaler2007}, Marthaler and Dykman calculated the tunnel-splitting between states comprising the ground state manifold and found that the tunnel splitting vanishes periodically as a function of the drive frequency. We massage our phase-space Hamiltonian function into a form resembling Equation 5 of Marthaler and Dykman \cite{marthaler2007} by rewriting \cref{eq:H-eff-ps} as
\begin{align}
\begin{split}
\label{eq:H-eff-ps-cl}
H = -\frac{K}{\lambda^2}\left[\left(\frac{x^2 + p^2}{2}\right)^2  - \frac{x^2}{2} \left( 1 + \frac{\Delta}{2 \epsilon_2}\right)   + \frac{p^2}{2} \left( 1 - \frac{\Delta }{2 \epsilon_2}\right)\right]
\end{split}
\end{align}
where their parameter $\mu = (\Delta + 2K)/2\epsilon_2$.
\subsection{WKB calculation of tunnel splitting for the ground state manifold of eq.~(1)}
\label{sec:WKB}
The expression for the tunnel splitting following the analysis in \cite{marthaler2006,marthaler2007} is given as
\color{black}
\begin{figure*}
\includegraphics{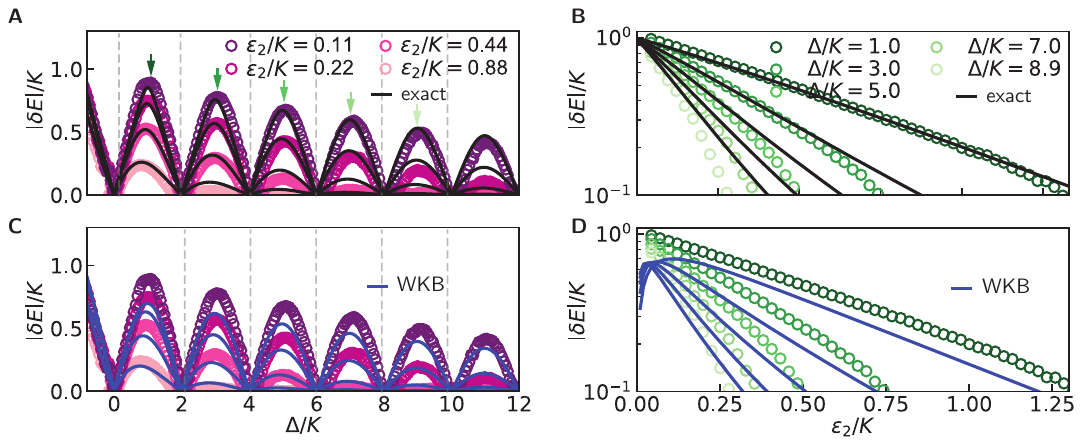}
\caption{Experimentally extracted tunneling amplitude in the ground state manifold as a function of $\Delta$ (\textbf{A} and \textbf{C}) and $\epsilon_2$ (\textbf{B} and \textbf{D}) compared to two different theoretical models. Dots correspond to extracted level splittings from dynamical measurements of the tunneling rate and correspond to the data presented in Figures 1C and 2 respectively. Solid lines in black in \textbf{A} and \textbf{B} are obtained via exact numerical diagonalization. Solid blue lines in \textbf{C} and \textbf{D} are obtained via a semi-classical WKB treatment developed by Marthaler and Dykman in \cite{marthaler2006,marthaler2007}. As expected, the semi-classical Hamiltonian model, in the domain of its validity $\Delta/K \gg 1$ and $\epsilon_2/K \sim 1$, agrees well with the measured data.}
\label{fig:WKB-diag}
\end{figure*}
\begin{align}
    \delta E = f \cos \theta \exp(-A),
\end{align}
where
\begin{align}
    \begin{split}
        f &=  2 \left(\frac{4 \epsilon_2}{K} \right)^2 \left(\frac{ K }{\pi (\Delta + 2K) }\right)^{1/2} \left(1 + \frac{(\Delta + 2K)}{2\epsilon_2} \right)^{5/4}  \\
        \theta &= \frac{\pi}{2} \left(\frac{(\Delta + 2K)}{K} - 1 \right)  \\
        A &= \frac{2\epsilon_2}{K} \left(\frac{(\Delta + 2K)}{2\epsilon_2} + 1\right)^{1/2}  - \frac{(\Delta + 2K)}{K} \log \left( \left(\frac{2\epsilon_2}{(\Delta + 2K)}\right)^{1/2} + \left(1 + \frac{2\epsilon_2}{(\Delta + 2K)}\right)^{1/2} \right),
    \end{split}
\end{align}
where, the above expression is only valid for $(\Delta + 2K)/K \gg 1$. There are two failure modes for the WKB approximation. The first condition corresponds to when $\Delta \lesssim K$, and the other is when $\epsilon_2/K \ll 1$. Note that WKB works remarkably well outside its domain of validity ($\epsilon_2/K < 1$). Compare to \Cref{fig:wf_lowest}, where the wavelength given by the oscillation period of the wavefunction is of the same magnitude as the potential variation set by the interwell distance. Note that we have applied the formula developed in \cite{marthaler2007} in a domain that lies beyond the parameter regime where it was produced and we find remarkable agreement with data. The comparison between measured tunneling amplitude and a WKB theory can be found in \Cref{fig:WKB-diag}.

\subsection{Action quantization via Einstein-Brillouin-Keller (EBK) method}
\label{sec:action}
In this section, we present the semiclassical method of obtaining the number of in-well states via action quantization, following the Einstein-Brillouin-Keller method, which generalizes the notion of Bohr orbits.

First, we introduce a polar-coordinate representation of \cref{eq:H-eff-ps-cl}, which exploits its radial symmetry, as
\begin{align}
\label{eq:H-eff-ps-cl-rad}
H_{\mathrm{cl}} = \frac{\Delta  r^2}{2} - \frac{K r^4}{4}  + \epsilon_2 r^2 \cos{2 \theta},
\end{align}
where $x = r \cos{\theta}$ and $p = r \sin{\theta}$, for $r \ge 0$ and $\theta \in [0, 2 \pi)$.

In a semiclassical treatment, a classical orbit $\mathcal C_j$ satisfying the following Einstein-Brillouin-Keller (EBK) quantization condition \cite{gutzwiller2013}:
\begin{align}
\label{eq:EBK-int}
    \int_{\mathcal C_j} d  x\,  dp = \hbar \left(N_j + \frac{\beta_j}{4} \right),
\end{align}
plays a special role. On the left hand side of \cref{eq:EBK-int}, the action integral corresponds to the area enclosed by the contour $\mathcal C_j$. On the right hand side of \cref{eq:EBK-int}, the non-negative integer $N_j \ge 0$ represents a quantum number and $\beta_j$ is called a Maslov index; it counts the number of caustics encountered by the contour $\mathcal C_j$. For an orbit in the Kerr-cat metapotential, we have $\beta_j = 2$.
Thus the condition in \Cref{eq:EBK-int} states that only those orbits whose enclosed area satisfy a condition given by non-negative integers $n_j$ and $\beta_j = 2$ correspond to allowed quantum orbits.

With this condition stated, one can ask a simple question: given a set of $\Delta$, $\epsilon_2$, how many in-well or bound states exist in the metapotential surface? This will be obtained by computing the number of allowed states at the separatrix, which separates bound and unbound states.

From the calculations detailed in \cref{area:double-node,area:kidneybean}, we find the number of bound states as
\begin{align}
\label{eq:Nstates}
    N \sim 
    \begin{cases}
    \frac{\Delta/K}{2} + \frac{\epsilon_2/K}{\pi} - \frac{1}{2} & -2 \epsilon_2 \le \Delta < 2\epsilon_2  \\
    \frac{\sqrt{8 \epsilon_2 \Delta}}{K \pi} - \frac{1}{2} & \Delta \ge 2 \epsilon_2.
    \end{cases}
\end{align}

We demonstrate in \cref{fig:wig-h} the value of the semi-classical action quantization condition in predicting the locality in phase space of even the excited states of the squeeze-driven Kerr oscillator. 

\subsubsection{Separatrix area in the double-node phase:  $-2 \epsilon_2 \le \Delta < 2\epsilon_2$}
\label{area:double-node}
In the double-node phase, the separatrix has a special name called the Bernoulli's lemniscate and its equation is given as
\begin{align}
    r^2 = \frac{2 \Delta}{K} + \frac{4 \epsilon_2}{K} \cos{2 \theta},
\end{align}

and $ -\theta_c \le \theta \le \theta_c $, where $\theta_c =  \frac{1}{2} \arccos{\frac{-\Delta}{2\epsilon_2}}.$ We compute the area of a half the lemniscate as

\begin{align}
\label{eq:area-double-node}
\begin{split}
    \int_{\mathcal C_j} d x\, p &= \frac{1}{2}\int_{-\theta_c}^{\theta_c} d\theta\, r^2 \\ 
    &=  \int_{0}^{\theta_c} d\theta\,  \frac{2 \Delta}{K} + \frac{4 \epsilon_2}{K} \cos{2 \theta} \\
    &= \frac{\Delta}{K} \arccos{\left(\frac{-\Delta}{2\epsilon_2}\right)} + \frac{2\epsilon_2}{K} \sqrt{1 - \left( \frac{\Delta}{2 \epsilon_2} \right)^2} \\
    &\sim \frac{\Delta}{ K} \left( \frac{\pi}{2} + \frac{\Delta}{2\epsilon_2} \right) + \frac{2 \epsilon_2}{K} \left(1 - \frac{1}{2} \left(\frac{\Delta}{2 \epsilon_2} \right)^2 \right) \qquad |\Delta/2\epsilon_2| \ll 1\\
    &=  \frac{\pi}{2} \frac{\Delta}{K} + 2 \frac{\epsilon_2}{K}
\end{split}
\end{align}
Note that for $\Delta = 0$, \cref{eq:area-double-node} reduces to $2 \epsilon_2/K$.

\subsubsection{Separatrix area in the triple-node phase: $\Delta \ge 2\epsilon_2$}
\label{area:kidneybean}
The separatrix in the triple-node phase is given as
\begin{align}
    r_{\pm}^2 = \frac{\Delta}{ K} + \frac{2 \epsilon_2}{K} \cos{2 \theta} \pm \frac{4 \epsilon_2 \cos \theta}{K} \sqrt{\frac{\Delta}{2 \epsilon_2} - \sin^2 \theta}
\end{align}
and $-\theta_c \le \theta \le \theta_c$, where $\theta_c = \frac{\pi}{2}$. When plotted, this separatrix carves a bean-like shape.

Remarkably, we find an exact analytic expression for the area of this surface as
\begin{align}
\begin{split}
     \int_{\mathcal C_j} d x\, p &= \frac{1}{2}\int_{-\theta_c}^{\theta_c} d\theta\, (r_+^2 - r_-^2) \\
     &= \int_{-\pi/2}^{\pi/2} d\theta\, \frac{4 \epsilon_2 \cos \theta}{K} \sqrt{\frac{\Delta}{2 \epsilon_2} - \sin^2 \theta} = \frac{4 \epsilon_2}{K} \int_0^1 dt\, \sqrt{\frac{\Delta}{2\epsilon_2} - t^2}\\
     &= \frac{4\epsilon_2}{K} \sqrt{\frac{\Delta}{2\epsilon_2} - 1} + \frac{2 \Delta}{K}\arcsin{\left(\sqrt{\frac{2\epsilon_2}{\Delta}} \right)} \\
     &\sim \frac{2 \sqrt{8 \epsilon_2 \Delta}}{K}, \qquad \Delta / 2\epsilon_2 \gg 1.
\end{split}
\end{align}

\section{Degeneracies in the squeeze-driven Kerr oscillator}
\label{sec:spectrum}

\subsection{Robustness of degeneracies}
\label{sec:deg}
The squeeze-driven Kerr oscillator we have engineered has the remarkable property: for $\Delta/K = 2m$, the first $m+1$ pairs of levels become decoupled from the rest of the oscillator's Hilbert space. Their eigenenergies and eigenstates become exactly solvable and present $m+1$ robust degeneracies in between states of different photon-number parity. Critically, note that the resonance condition for these degeneracies is independent of the value of the squeezing drive amplitude $\epsilon_2$.

First, to show this, we begin by considering the squeezing drive as a perturbation to the Kerr oscillator described by the Hamiltonian $\hat H_K/\hbar = \Delta \hat a ^{\dagger } \hat a - K \hat a ^{\dagger 2} \hat a^2$ which is exactly solvable: its eigenstates are Fock states $|n\rangle$ and their energies are $E^{(0)}_n  = \Delta n-K n(n-1) $, which, as a function of $\Delta$, are lines with integer slope that we plot in the top row of \Cref{fig:spectrum}\textbf{A}. The even($n$)-odd($l$) degeneracies read $E_n = E_l$ and imply $\Delta/K = 2m$ where $m=(n+l-1)/2\geq 0$ is any nonnegative integer. In the second row, we plot the transition spectrum with respect to the ground state at $\epsilon_2 = 0$, which, due to the choice of rotating frame, corresponds to the \textit{highest} energy eigenstate. This is the directly experimentally observable \textit{transition} spectrum from the \textit{ground} state. We further note that the ground state changes with $\Delta$; remarkably, for $\epsilon_2 = 0$, at $\Delta/K = 2m$, the ground state is $(|m \rangle + |m+1\rangle)/\sqrt{2}$. This special property of the squeeze-driven Kerr oscillator has technological applications \cite{zhang2017}. In the following rows, we plot the transition spectrum for increasing values of squeezing drive amplitude $\epsilon_2$.

Indeed, it is clear that the squeezing drive renormalizes the energies of the Kerr oscillator. Level crossings of the Kerr oscillator with different parity remain exact crossings in the presence of the squeezing drive, since the interaction preserves parity. However, the remarkable feature is that these crossings are locked to where $\Delta$ equals an even multiple of $K$. In the following text, we justify this property, first via a perturbative and then provide a to-all-order proof.

\subsubsection{Perturbative analysis of degeneracies}
To first order in perturbation theory, we see that this even and odd Fock states remain decoupled (energy level crossings) under the parity conserving squeezing drive: $E_n^{(1)} = \langle n|(\hat a^{\dagger 2} + \hat a^2)|n+1\rangle = 0.$ The condition for crossings of consecutive levels with different parity ($E_n = E_{n+2}$) reads instead $\Delta/K = (2n+1)$. To first order in perturbation theory, the avoided crossing amplitude is $E_n^{(1)}=\epsilon_2 \sqrt{(n+1)(n+2)}.$

\begin{figure*}
\includegraphics{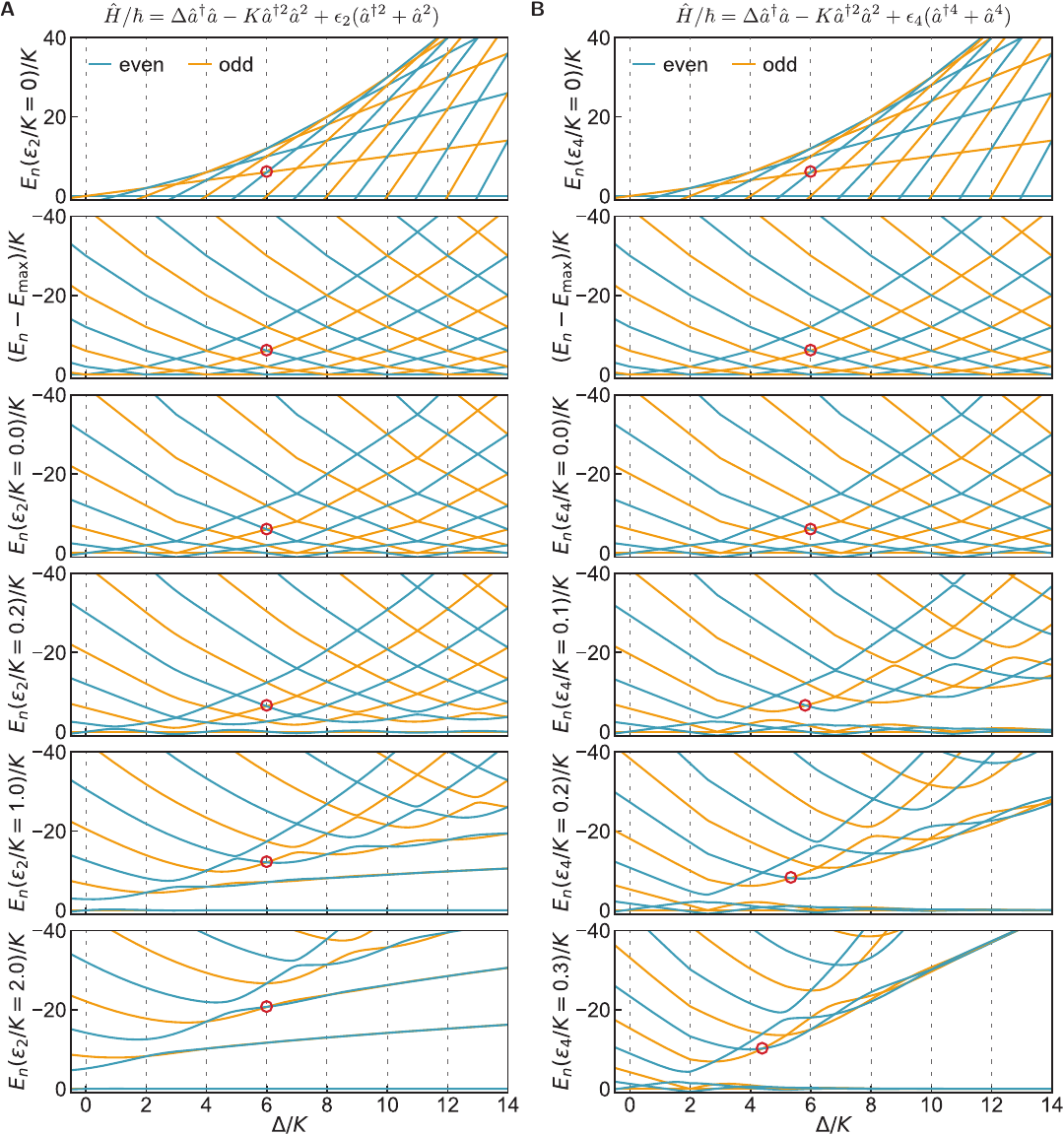}
\caption{\textbf{Robustness of degeneracies in the squeeze-driven Kerr oscillator. }  Spectrum of \textbf{A} Eq. (1) and \textbf{B} $\hat{H} = \Delta \hat{a}^\dagger \hat{a} - K \hat{a}^{\dagger 2} \hat{a}^2 + \epsilon_4 (\hat{a}^{\dagger 4} + \hat{a}^4)$ as a function of $\Delta/K$ for different values of $\epsilon_2/K$ and  $\epsilon_4/K$ respectively. Dashed lines mark $\Delta/K$ corresponding to even integers. Left panel indicates that even for non-perturbative values of $\epsilon_2/K$, the locations of crossings of even (blue) and odd (orange) parity eigenstates occur at even values of $\Delta/K$. Right panel indicates that even for the parity preserving perturbation controlled by $\epsilon_4/K$, the locations of the crossings of even and odd parity states get renormalized. Red circle tracks one such crossing.}
\label{fig:spectrum}
\end{figure*}

As a next approximation to the problem, we see the robustness of the crossings of consecutive levels with different parity (at $\Delta = 2nK$) by computing the second order correction to the $n$th energy levels $ E^{(2)}_n$ and comparing it to the correction for the $(n+1)$th energy level
$$ E^{(2)}_{n+1} = \epsilon_2^2\left(\frac{(n+3)(n+2)}{-2\Delta + 2K(2n+3)} + \frac{(n+1)n}{2\Delta - 2K(2n+1)}\right),$$
to find that $E^{(2)}_n =E^{(2)}_{n+1}$ for $\Delta/K = 2n$. This robustness can be seen in \Cref{fig:spectrum}\textbf{A} (all panels), where we see that the crossing shifts in energy but remains locked to $\Delta/K$ equal to even non-negative integers. The perturbation theory argument is easily generalized to non-consecutive level crossings and anti-crossings to this order. A similar perturbative argument was made in \cite{zhang2017}.
\subsubsection{Non-perturbative analysis of degeneracies}
To prove that the location of the degeneracies in $\Delta$ is independent of the squeezing drive amplitude to all orders we observe that we can write the Hamiltonian in Eq.~(1) as
\begin{align}
    \label{eq:Michelsfactorization}
\hat{H} = \lambda_1 (\hat a^{\dagger 2}-\alpha^2)(\hat a^{ 2}-\alpha^2) + \lambda_2 (\hat a^{ 2}-\alpha^2)(\hat a^{ \dagger 2}-\alpha^2),
\end{align}
where, for $\Delta/K =2m$ ($m$ non-negative integer), we have $\lambda_1 = -K(1+m/2)$, $\lambda_2 = mK/2$ and $\alpha=\pm\sqrt{\epsilon_2/K}$, and which is a generalization of the factorization condition proposed in \cite{puri2017} for $\Delta=0$.
We next consider the displaced Hamiltonian $\hat H^+ = \hat D(+\alpha) \hat H \hat D^{\dagger}(+\alpha)$, which brings one of the wells to the origin of phase space. In this frame, the Hamiltonian operator can be written as\footnote{Note that, without specializing $\Delta$, one can directly write from Eq.~(1) in the main text, or equivalently from  \cref{eq:Michelsfactorization}: $\hat{H}^+ = -K\left(\hat a^{\dagger 2}\hat a^{ 2} +(4\alpha^2+\Delta/K)\hat a^{\dagger }\hat a\right)-2K\alpha [\hat a^{\dagger }\hat a - (\Delta/2K+1)]\hat a^{\dagger }-2K\alpha [\hat a^{\dagger }\hat a - \Delta/2K]\hat a$. From this expression one can directly derive the sub-space decoupling condition to be $\Delta/K = 2m$, in an exact manner, without relying in perturbative calculation or any previous knowledge existence of the resonance. The independence of the sub-space decoupling condition with respect to $\epsilon_2$ is explicit.}

\begin{align*}
    \hat{H}^+ = &-K\left(\hat a^{\dagger 2}\hat a^{ 2} +(4\alpha^2+2m)\hat a^{\dagger }\hat a\right)\\
    &-2K\alpha [\hat a^{\dagger }\hat a - (m+1)]\hat a^{\dagger }\\
    &-2K\alpha [\hat a^{\dagger }\hat a - m]\hat a.
\end{align*}

While the first line is number conserving, the next two lines couple only consecutive Fock states. In matrix form, it is tridiagonal in the Fock basis $|n\rangle$.
By examining the square brackets in the above expression, we see that the off-diagonal elements are exactly zero for $n=m$ and $n=m+1$. Thus, the first $m+1$ states decouple from the rest of the oscillator's Hilbert space. The finite matrix is Hermitian, negative-semidefinite, and tridiagonal so it is exactly diagonalizable.
Finally, we note that in phase space, a displacement of the metapotential surface, which is mirror-symmetric about $x=0$, is identical to an opposite displacement composed with a rotation of $180^{\circ}$ around the origin. Since the photon-number parity operator $\hat \Pi = e^{i\pi\hat a ^{\dagger} \hat a}$ commutes with the Hamiltonian ($[\hat \Pi, \hat H]=0$) the rotation is a symmetry of the system. Specifically; $ \hat H^- = \hat D(-\alpha) \hat H \hat D^{\dagger}(-\alpha)\Rightarrow \hat \Pi \hat H^- \hat \Pi =  \hat H^+$. We thus have two sets of equivalent\footnote{Note that for the off-diagonal elements the parity transformation produces a minus sign ($\hat \Pi |n\rangle\langle n\pm1|\hat \Pi =-|n\rangle\langle n\pm1|$) that manifests in $\alpha\rightarrow-\alpha$: $H^+_{n,n\pm 1} = -H^-_{n,n\pm 1}$. This leaves the finite characteristic polynomial invariant.} $m+1$ exactly solvable eigenenergies, and $2(m+1)$ linearly independent equations \footnote{The elements of the finite set of eigenvectors of $\hat H^{+}$, $\{|\phi^{+}_{k\leq m+1}\rangle\}_k$, are linearly independent from the finite set of eigenvectors of $\hat H^{-}$, $\{|\phi^{-}_{k\leq m+1}\rangle\}_k$, since they are spanned by the first $m+1$ displaced Fock states in different directions ($\pm$, see text). Indeed, any Fock state $i$ that is displaced has support in all (undisplaced) Fock states $j$'s \cite{Cahill1969}: for $j>i$ the formula reads $|\langle j|\hat D(\pm2\alpha)|i\rangle|=\left(\frac{i!}{j!}\right)^{1/2}|2\alpha|^{j-i}e^{-2|\alpha|^2}|L_i^{(j-i)}(4|\alpha|^2)|>0$, where $L_i^{(j-i)}$ is an associated Laguerre polynomial (note that the matrix element tends to zero rapidly as $|\alpha|$, or equivalently $|\epsilon_2|$, grows, while the decoupled subspace condition, and ultimately the proof itself, is independent of these values for as long as they are non zero. If $\epsilon_2=0$, the proof is trivial and is given in the previous subsection). In other words, the linear independence manifests here explicitly in that $|\phi^{\pm}_{k\leq m+1}\rangle$ have no defined parity, yet $|\phi^{+}_{k\leq m+1}\rangle=\hat \Pi|\phi^{-}_{k\leq m+1}\rangle$ (note that $[\hat \Pi, \hat H^{\pm}]\neq 0$). Ultimately, $|\langle \phi^{-}_{k\leq m+1}|\hat D^{\dagger}(-\alpha)\hat D(+\alpha)|\phi^{+}_{k\leq m+1}\rangle|=|\langle \phi^{+}_{k\leq m+1}|\hat D(2\alpha)|\phi^{+}_{k\leq m+1}\rangle|<1$ if $|\alpha|>0$.}, which imply the existence of $m+1$ degeneracies in the spectrum for $\Delta/K = 2m$. The $2(m+1)$ eigenstates $|\psi^{\pm}_{k\leq m+1}\rangle = \hat D(\pm\alpha)|\phi^{\pm}_{k\leq m+1}\rangle$ of $\hat H$, where $|\phi^{-}_{k\leq m+1}\rangle = \hat \Pi |\phi^{+}_{k\leq m+1}\rangle$ and $\hat H^+|\phi^{+}_{k\leq m+1}\rangle = E_{k\leq m+1}|\phi^{+}_{k\leq m+1}\rangle$, found in this way are not orthogonal, but thanks to the two-fold degeneracy condition we can take the superposition of the right ($+$) and left ($-$) $k$th displaced state to get an orthogonal basis in each of the $m+1$ two-fold degenerate sub-spaces: $|\mathcal C^{\pm}_{k\leq m+1}\rangle \propto D(+\alpha)|\phi^{+}_{k\leq m+1}\rangle \pm D(-\alpha)|\phi^{-}_{k\leq m+1}\rangle$. These $2(m+1)$ pairwise-degenerate eigenstates of energy are also eigenstates of parity\footnote{Specifically $\hat \Pi |\mathcal C^{\pm}_{k\leq m+1}\rangle = \pm |\mathcal C^{\pm}_{k\leq m+1}\rangle$ and are thus orthogonal. We used $\hat D(+\alpha) \hat \Pi = \hat \Pi \hat D(-\alpha)$.}. In this work we name these pairs of degenerate states the $\Delta$-cats.


Note, that the robustness of the resonance condition is a peculiar symmetry property of the squeeze-driven Kerr oscillator and not a property of generic Kerr parametric oscillators. The existence of this robust degeneracies begs the question: what are the hidden symmetries associated with these degeneracies, if any? We show in \Cref{fig:spectrum}\textbf{B}, as an example, the spectrum of $\hat{H} = \Delta \hat{a}^\dagger \hat{a} - K \hat{a}^{\dagger 2} \hat{a}^2 + \epsilon_4 (\hat{a}^{\dagger 4} + \hat{a}^4),$ where the location in $\Delta$ of the super-parity resonances depend on the value of the parametric drive amplitude $\epsilon_4$. Note, also, that even if the multilevel resonances in \Cref{fig:spectrum}\textbf{B} are displaced with the value of the parametric drive amplitude (red circles), they are locked together to a running resonance condition: the point of exact solvability is changed by the drive. The phenomenon corresponds to deep symmetries \cite{iachello1979interacting,iachello2015}  of these type of, as of now, engineerable bosonic Hamiltonians and will be discussed in detail in a separate publication.

\section{Modeling the measured transverse relaxation lifetime $T_X$}
\label{model}
\begin{figure*}
\includegraphics{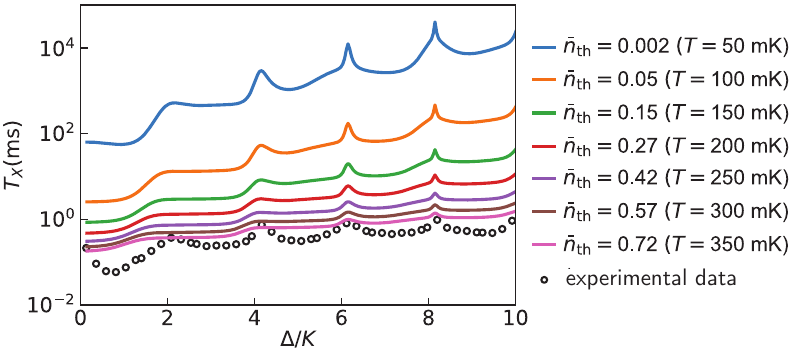}
\caption{\textcolor{black}{\textbf{Lindblad simulations of $T_X$ as a function of $\Delta$ for different thermal populations, corresponding to \cref{eq:ord-me}.} Black dots correspond to experimental data presented in Fig. 3 \textbf{G} in the main text. The value of $\kappa$ has been taken as $\kappa = 1/T_{1} =  1/20~$\textmu s$^{-1}$ and the value of $\epsilon_2$ has been chosen as $\epsilon_2/K = 2.17$ to match the experimental data. The solid curves take the experimentally observed ac Stark shift into account. An ordinary Lindbladian at non-zero temperature is insufficient to predict the experimental data. Beyond-RWA effects may be important to consider \cite{venkatraman2022_2}. See also \cite{Dykman1998}.}}
\label{fig:Lind-sims-vsDelta}
\end{figure*}

\begin{figure*}
\includegraphics{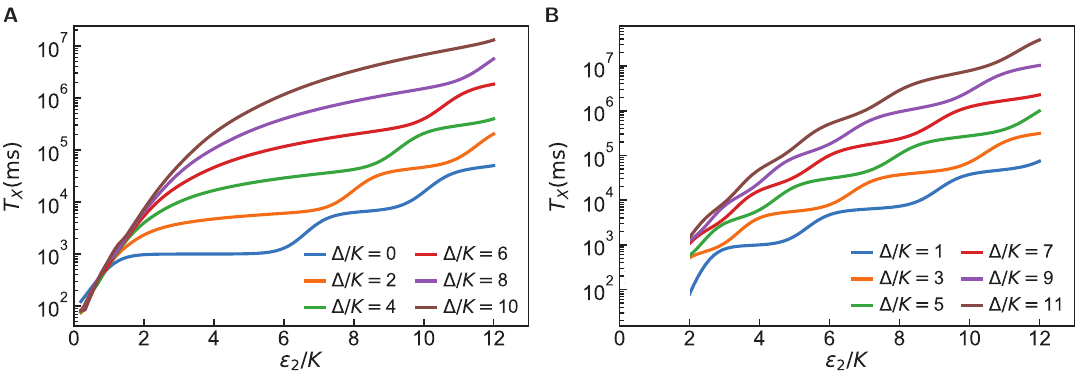}
\caption{\textcolor{black}{\textbf{Ordinary Lindblad simulations of $T_X$ as a function of $\epsilon_2/K$ for different values of $\Delta/K$,, corresponding to \cref{eq:ord-me}.} For both \textbf{A} and \textbf{B}, the value of $\kappa/K = 1/50$ and $\bar{n}_{\mathrm{th}} = 0.05$. In \textbf{B} for $\epsilon_2/K<2$ the lifetime is limited by ground state tunneling and is this not well captured by our simplified method.}}
\label{fig:Lind-sims-vse2}
\end{figure*}

To model the transverse relaxation lifetime measurements $T_X$ of the Kerr-cat qubit, which we also refer to as the well-switching lifetime of the Kerr-cat system, we use a standard Lindblad master equation as: 

\begin{align}
\label{eq:ord-me}
\partial_t \hat \rho = \frac{1}{i\hbar}[\hat H,\hat \rho]+ \kappa (1 + \bar{n}_{\mathrm{th}}) \mathcal D[\hat a] \hat{\rho} + \kappa \bar{n}_{\mathrm{th}} \mathcal D[\hat a^\dagger] \hat{\rho},
\end{align}
where $\hat \rho$ describes the state of the system, $\bar{n}_{\mathrm{th}} = 1/(\exp(\hbar \omega_a/ k_B T ) - 1)$ corresponds to the temperature of the environment and $\kappa$ corresponds to the coupling between system and environment. The Hamiltonian $\hat H$ is given by \cref{eq:H-eff} and the dissipator $\mathcal D$ of the operator $\hat{O}$ is given by $\mathcal D[\hat O]\bullet := \hat O\bullet\hat O^\dagger -(\hat O^\dagger \hat O\bullet + \bullet\hat O^\dagger \hat O)/2$. In \cref{eq:ord-me}, these operators correspond to single photon loss $\mathcal{D}[\hat{a}]$ and gain $\mathcal{D}[\hat{a}^{\dagger}]$ \cite{carmichael1999,carmichael2009,breuer2002}. In \Cref{fig:Lind-sims-vsDelta}, we compare the data presented in Figure 3 of the main text with the lifetime extracted from \cref{eq:ord-me} for different values of $n_{\mathrm{th}}$. The value of $\kappa$ has been set to $\kappa = 1/T_{1} =  1/20~$\textmu s$^{-1}$. The current model seems insufficient to accurately predict the observations and more research is needed to understand the decoherence of nonlinear driven systems (see, for example, \cite{venkatraman2022_2}). \Cref{fig:Lind-sims-vsDelta} emphasizes the need for further measurements and a detailed modeling of possible noise sources affecting particularly driven qubits. See also the note at the end of \cref{sec:tutorial}. We also present, in \Cref{fig:Lind-sims-vse2}, the expected $T_X$ as a function of $\epsilon_2/K$ for different values of $\Delta$. This plot indicates that a $\Delta$-Kerr-cat, in general, gives larger $T_X$ lifetimes than a Kerr-cat ($\Delta = 0$).

\color{black}

\begin{widetext}

\pagebreak

\section{Tutorial on the phase space formulation of quantum mechanics}
 \label{sec:tutorial}
A full quantum mechanical treatment can be developed in phase space without incurring in any semiclassical approximations \cite{moyal1949,groenewold1946,curtright2013}. For the sake of completeness, here we provide an overview on the mapping from operator-valued Hilbert space to quantum phase space and a few elemental techniques and identities. We focus here on Wigner phase space, and showcase that the Wigner transform is more than a visualization tool for states. We note that our treatment can be equivalently extended to other phase space formulations \cite{raimond2006,puri2001,curtright2013,hillery1984,Xiao2022}.

\section*{From operator Hilbert space to Wigner phase space (and back)}

The Wigner transform \cite{wigner1932} of the density matrix $\hat \rho$ is the Wigner function $W(X,P)$, where $X$ and $P$ are standard phase space coordinates (not operators) with dimensions of position and momentum (see \cref{sec:notation} for notation). We write this as

$$\mathfrak{W} \{ \hat \rho \}=W(X,P).$$

Let us remind the reader of some crucial properties of the Wigner function. We have
\begin{align}
    \int\!\!\!\!\int dX dP \, W(X, P ) = 1,
\end{align}
where each integral runs from $-\infty$ to $\infty$ and we suppress the limits in the following text for simplicity.
For a pure state, we further have
\begin{align}
    h \int\!\!\!\!\int dX dP \, W(X, P )^2 = 1,
\end{align}
where $h = 2\pi \hbar$. 

In general, we have
\begin{align}
    0 \le h^{n-1} \int\!\!\!\!\int dX dP \, W(X, P )^n \le 1,
\end{align}
which corresponds to the positivity of the density matrix.

Likewise, for a generic operator $\hat F$, we introduce the phase space function $F(X,P)  = \mathfrak{W} \{ \hat F \}.$

In this framework, the average value of an Hermitian operator $\hat{F}$ can be written as
\begin{align}
   \langle \hat F \rangle = \int\!\!\!\!\int dX dP\, F(X, P) W(X, P).
\end{align}

The transformation $\mathfrak W$ is invertible as appreciated by Groenewold \cite{groenewold1946}

$$\mathfrak{W}^{-1}\{W(X,P)\}=\hat \rho.$$ 
The inverse transformation $\mathfrak{W}^{-1}$ is know as the Weyl transformation \cite{weyl1931}.

In general, the Weyl transformation is
\begin{align}
    \hat\rho = \mathfrak{W}^{-1}\{ W \} = \frac{1}{h} \int\!\!\!\!\int\!\!\!\!\int\!\!\!\!\int dX dP dk dl \, W(X, P) e^{\frac{i}{\hbar} (k (\hat{X} - X) + l(\hat{P} - P))},
\end{align}
where the characteristic function $C(l,k)$ defined as
\begin{align}
    C(l,k) = \int\!\!\!\!\int dX dP \, e^{-\frac{i}{\hbar} (kX + lP)} W(X, P) ,
\end{align}
is the Fourier transform of the Wigner function and $C$ is dimensionless.

Another useful formula is
\begin{equation}
\label{eq:WtranformIntegralW}
W(X,P)=\frac{1}{h} \int d q \, e^{-i qP / \hbar} \langle X+q / 2|\hat{\rho}| X-q / 2\rangle,
\end{equation}
where $\hat{\rho}$ is to be understood in the continuous position basis and therefore has the dimension of [1/position].

We now review simple operational rules to go from operator space to phase space functions and back without performing cumbersome integrals.

The Wigner and Weyl transformation take a particularly simple form for binomial expansions

$$\mathfrak{W}\{(\alpha \hat X+\beta \hat P)^n\}=(\alpha X + \beta P)^n.$$
$$\mathfrak{W}^{-1} \{(\alpha X + \beta P)^n \}=(\alpha \hat X+\beta \hat P)^n.$$

For non-symmetric expressions, the Wigner transform can be evaluated via a non-commutative Wigner phase space product, the celebrated Groenwold's star product.

\subsection{An introduction to the star product}

We introduce the star product as

\begin{equation}
\label{eq:star1}
    \mathfrak{W}(\hat F\hat G)=\mathfrak{W}(\hat F)\star\mathfrak{W}(\hat G) = F(X,P)\star G(X,P),
\end{equation} 
defined as (the exponential of the Poisson bracket):
$$F \star G=\sum_{n=0}^{\infty} \sum_{k=0}^{n} \frac{(-1)^{k}}{n !}\left(\frac{i \hbar}{2}\right)^{n}\binom{n}{k} \partial_{P}^{k} \partial_{X}^{n-k} F \times \partial_{P}^{n-k} \partial_{X}^{k} G $$

\begin{equation}
\label{eq:star2}
    \equiv F \exp \left(\frac{i \hbar}{2}\left(\overleftarrow{\partial}_{X} \overrightarrow{\partial}_{P}-\overleftarrow{\partial}_{P} \overrightarrow{\partial}_{X}\right)\right) G
\end{equation} 
$$= FG + \frac{i\hbar}{2}\{F,G\}+\cdots$$

Here $F\overleftarrow{\partial}_{X}G=(\partial_X F)G$ and $F\overrightarrow{\partial}_{X}G=F(\partial_X G)$, and we have introduced the Poisson bracket $\{F,G\} = \partial_X F\partial_P G-\partial_P F\partial_X G$. The star product can also be conveniently expressed in terms of complex-coordinates $a$ and $a^*$ as

$$F \star G\equiv F\exp \left(-\frac{1}{2}\left(\overleftarrow{\partial}_{a^*} \overrightarrow{\partial}_{a}-\overleftarrow{\partial}_{a} \overrightarrow{\partial}_{a^*}\right)\right) G.$$

It generalizes to a system of many particles (or many modes) as
$$F \star G=F \exp \left(\frac{i \hbar}{2}\sum_j\left(\overleftarrow{\partial}_{X_j} \overrightarrow{\partial}_{P_j}-\overleftarrow{\partial}_{P_j} \overrightarrow{\partial}_{X_j}\right)\right) G.$$

In Fourier space the star product becomes a phase factor: $\star\rightarrow e^{i\frac{\hbar}{2}(k_X k'_P-k'_X k_P)}$ \cite{zachos2000}. This phase corresponds to an oriented area in reciprocal phase space. This is the simplest manifestation of the noncommutativity of the algebra of quantum mechanics in phase space.

Remarkably, the scalar product associated with the star product is the usual integral in phase space. For phase space functions in the Wigner representation $F$ and $G$, we have
\begin{align}
 \int\!\!\!\!\int dX dP\, F(X, P) \star G(X, P) = \int\!\!\!\!\int dX dP\, F(X, P) G(X, P).
 \end{align}
 Note however that in general for any $F(X, P)$, $G(X,P)$, and $H(X,P)$,
 \begin{align}
 \int\!\!\!\!\int dX dP\, F(X, P) \star G(X, P) \star H(X,P) \neq \int\!\!\!\!\int dX dP\, F(X, P) G(X, P) H(X,P).
 \end{align}

For non-symmetric expressions in $\hat X$ and $\hat P$, the above formulae can be employed to evaluate the Wigner transform. For example

$$\mathfrak{W} \{\hat{X} \hat{P} \hat{P}\} = XP^2 + i \hbar P$$
$$\mathfrak{W} \{ \hat{P} \hat{P} \hat{X}\} = XP^2 - i \hbar P$$
$$\mathfrak{W} \{ \hat{P} \hat{X} \hat{P}\} = XP^2.$$

We evaluate the Weyl transform of asymmetric expressions by symmetrizing it and replacing phase space functions by their corresponding operators. For example
$$\mathfrak{W}^{-1} \{X P^2\} = \frac{1}{3}(\hat{X} \hat{P} \hat{P} + \hat{P} \hat{X} \hat{P}  + \hat{P} \hat{P} \hat{X}).$$

To find the Weyl transform of a high-degree polynomial of $X$ and $P$, the Weyl-symmetrized form might be too tedious and McCoy \cite{mccoy1932} provided a shortcut to obtain polynomial expressions in the phase space representation. We review McCoy's formula in the next section.

\section*{The McCoy formula for obtaining ordered operators from phase space functions}

While a fully-symmetrized representation is usually inconvenient for polynomials of large degree, McCoy derived a set of formulae \cite{mccoy1932}, each corresponding to a different representation of a Weyl transform. Here, we present two of them that yield operators that privilege the ordering of $\hat{X}$ (or $\hat{P}$).

Consider a phase space function $F(X,P)$. Its operator-valued correspondent $\hat F$ in normal order with respect to $X$ is given by the McCoy formula \cite{mccoy1932} that reads:

$$\mathcal F(X,P) = e^{-i\frac{\hbar}{2}\partial_{X}\partial_{P}}F(X,P)$$
$$\mathcal F(X,P)=F(X,P) - \frac{i\hbar}{2}\partial_{X}\partial_{P}F(X,P) - \frac{1}{2!}\frac{\hbar^2}{2^2}\partial^2_{X}\partial^2_{P}F(X,P)+\cdots$$

$$\hat F  = (\mathcal N _X \mathcal F)|_{(\hat X, \hat P)},$$

The functional (operator over real functions) $\mathcal N_X$ is carried out by writing its arguments with $X$ factors (or $P$ factors as indicated by the subindex of $\mathcal N$) to the left in each term and replace $X, P$ with $\hat{X}, \hat{P}$ respectively. 
For example, if $ F =XP$, we have $\mathcal F=XP-i\frac{\hbar}{2}$ which gives the correct and now ordered Hermitian expression for the operator $\hat F = \hat X\hat P -i\frac{\hbar}{2}=\frac{\hat X \hat P+\hat P \hat X}{2}$.

The inverse transform, is simply given $F(X,P) = e^{i\frac{\hbar}{2}\partial_{X}\partial_{P}}\mathcal F(X,P)$.

In terms of complex coordinates, $a=\frac{1}{\sqrt{2}}(x+ip)$ we adapt McCoy's formula \cite{mccoy1932}: 

$$\mathcal F(a,a^*)=e^{\frac{1}{2}\partial_a\partial_{a^*}} F(a,a^*)$$
$$\hat F  = (\mathcal N _{a^*} \mathcal F)|_{(\hat a, \hat a^{\dagger})}$$
to get the normal ordered (with respect to $a^*$) result. For example, one has classically that $\frac{1}{2}(x^2+p^2) = a^*a$. The correct quantization reads $F=aa^*\rightarrow  \mathcal F=aa^* + 1/2\rightarrow  \hat F=\hat a^{\dagger} \hat a + 1/2.$

\subsubsection*{Application to our Hamiltonian}
If the Wigner phase space Kerr Hamiltonian reads $H= \Delta a^{*}a - Ka^{*2}a^2$ the corresponding operator is 

$$F=a^{*2}a^2\rightarrow  \mathcal{F}=a^{*2}a^2 + 2  a^{*}a + \frac{1}{2}\rightarrow  \hat F=\hat a^{\dagger2} \hat a^2 + 2 \hat a^{\dagger} \hat a + \frac{1}{2},$$

$$\hat H/\hbar = (\Delta - 2K)\hat a^{\dagger} \hat a - K\hat a^{\dagger2} \hat a^2,$$

where the oscillator frequency is renormalized by $2K$. This is the Lamb shift, and its origin is in the non commutativity of $\hat{a}$ and $\hat{a}^{\dagger}$, i.e. the vacuum fluctuations.

\section*{Groenewold's theorem}

Note that $\mathfrak{W}\left\{\frac{1}{i\hbar}[\hat F, \hat G ]\right\} = \{\!\!\{\mathfrak{W}(\hat F) ,\mathfrak{W}(\hat G)\}\!\!\}\neq \{\mathfrak{W}(\hat F) ,\mathfrak{W}(\hat G)\}.$ The quantum commutators do not correspond to the Poisson brackets: the theorem \cite{groenewold1946} states that such a mapping does not exist. We provide a practical consequence of the implications of this theorem to quantum Hamiltonian engineering in Appendix B of \cite{venkatraman2022}.

\section*{Dynamics of the Wigner function: the Moyal equation}

The von-Neumann equation $\partial_t \hat\rho =  \frac{1}{i\hbar}[\hat H,\hat \rho]$ (the density-operator version of the Schr\"odinger equation) transforms as

$$\partial_t W = \frac{1}{i\hbar}(H\star W - W\star H),$$

$$\partial_t W = \{\!\!\{H,W\}\!\!\}.$$

Here $ H(X,P) = \mathfrak{W}(\hat H)$ is the Hamiltonian function and we have introduced the Moyal bracket notation \cite{moyal1949}. We refer the reader to \cite{case2008} for a derivation of the equation of motion of the Wigner function from Sch\"odinger's equation for the wavefunction without referring to the star product.

The exponential notation of the star product induces the name ``Moyal sine bracket'' since it can be written as
$$\partial_t W = H\frac{2}{\hbar}\sin\left(\frac{ \hbar}{2}\left(\overleftarrow{\partial}_{X} \overrightarrow{\partial}_{P}-\overleftarrow{\partial}_{P} \overrightarrow{\partial}_{X}\right)\right)W.$$

Note that the Moyal equation is identical to Liouville equation plus quantum corrections coming from the expansion of the sine to higher orders of $\hbar$.

$$\partial_t W = \{H,W\} + \mathcal O(\hbar^2).$$

Interestingly, there is no corrections to $\mathcal{O}(\hbar)$. Importantly, the quantum corrections are proportional to $\hbar^2$ and to the nonlinear terms in the Hamiltonian. For quadratic Hamiltonians, all the quantum corrections vanish: the higher-order derivatives exterminate low-order polynomials (see the Appendix of \cite{frattini2022}). Specifically, Gaussian transformations, i.e., those generated by quadratic Hamiltonians in the phase space coordinates, are classical in the sense that they are ruled by only the Poisson bracket. Thus, they would not develop negativities in the Wigner distribution if none would be present at the beginning.

\section*{Phase space formulation for open quantum systems}

So far, we have only discussed the phase space formulation for closed quantum systems. Indeed, one can extend the treatment to open systems as we demonstrate below. The Lindblad equation for single photon loss is given by 

\begin{align}
\label{eq:VN}
    \partial_t \hat \rho = \frac{1}{i\hbar}[\hat H,\hat \rho]+\kappa \hat a \hat \rho\hat a^{\dagger}-\frac{\kappa}{2}(\hat a^{\dagger}\hat a \hat \rho+\hat \rho \hat a^{ \dagger}\hat a).
\end{align}
Using \cref{eq:star1} and \cref{eq:star2}, we get the phase space formulation of \cref{eq:VN} as
\begin{align*}
\mathfrak{W}\{\partial_t\hat \rho\}=\partial_t W,
\end{align*}
\begin{align*}
\mathfrak{W}\left\{\frac{1}{i\hbar}[\hat H, \hat \rho ]\right\} &= \{\!\!\{\mathfrak{W}(\hat H) ,\mathfrak{W}(\hat \rho)\}\!\!\} \\
&= \{\!\!\{H ,W\}\!\!\}
\end{align*}
\begin{align*}
\mathfrak{W}\{\hat a \hat \rho\hat a^{\dagger}\}&=a\star W\star a^*\\
&=aWa^*+\frac{1}{2}\partial_aW+\frac{1}{2}(\partial_{a^*}(Wa^*)+\frac{1}{2}\partial^2_{a a^*}W)
\end{align*}
\begin{align*}
\mathfrak{W}\{\hat a^{\dagger}\hat a \hat \rho\}&=\left( a^{*} a -\frac{1}{2}\right)\star W\\
&=aWa^*-\frac{1}{2}W+\frac{1}{2}(a^*\partial_{a^*}W-a\partial_aW)-\frac{1}{4}\partial^2_{a a^*}W
\end{align*}
\begin{align*}
\mathfrak{W}\{ \hat \rho\hat a^{\dagger}\hat a\}&=W\star \left( a^{*} a -\frac{1}{2}\right)\\
&=aWa^*-\frac{1}{2}W-\frac{1}{2}(a^*\partial_{a^*}W-a\partial_aW)-\frac{1}{4}\partial^2_{a a^*}W
\end{align*}

Gathering all terms one directly gets
$$
\partial_t W = \{\!\!\{H,W\}\!\!\}+\frac{\kappa}{2}\left(\partial^2_{a a^*}+\partial_a a +\partial_{a^{*}} a^{*}\right) W.
$$
It is convenient to translate the above to $x, p$ space
\begin{align}
\label{eq:diff}
\partial_t W = \{\!\!\{H,W\}\!\!\}+\frac{\kappa}{2}\left(\partial^2_x+\partial^2_p+\partial_xx+\partial_pp\right) W.
\end{align}
By expressing the equation in $x, p$ space in \cref{eq:diff}, the diffusion terms $\propto (\partial^2_x+\partial^2_p)$ and the drag terms $\propto (\partial_xx+\partial_pp$) associated to the fluctuation and the dissipation become evident. Note, that the Moyal sine bracket has only odd derivatives: the diffusion ($\partial^2_x+\partial^2_p$) cannot be canceled by Hamiltonian dynamics.

For finite temperature $\bar{n}_{\mathrm{th}}$, the Lindblad master equation is
\begin{align}
\label{eq:Lindblad-op}
   \partial_t \hat \rho = \frac{1}{i\hbar}[\hat H,\hat \rho]+ \kappa (1 + \bar{n}_{\mathrm{th}}) \mathcal D[\hat a] \hat{\rho} + \kappa \bar{n}_{\mathrm{th}} \mathcal D[\hat a^{\dagger}] \hat{\rho},
\end{align}
where the dissipator $\mathcal D$ of the operator $\hat{O}$ is given by $\mathcal D[\hat O]\bullet := \hat O\bullet\hat O^\dagger -(\hat O^\dagger \hat O\bullet + \bullet\hat O^\dagger \hat O)/2$.

It is straightforward to show that in the phase space formulation,  \cref{eq:Lindblad-op} reads
\begin{align}
\begin{split}
\partial_t W &=\{\!\!\{H, W\}\!\!\} + \frac{\kappa}{2} \left(    \partial_a a  + \partial_{a^*}   a^*  \right) W + \kappa \left(\frac{1}{2} + \bar{n}_{\mathrm{th}}\right)  \partial^2_{a^* a} W,
\end{split}
\end{align}
which reads in $x, p$ space as
\begin{align}
\label{eq:FP-Q}
\begin{split}
\partial_t W &=\{\!\!\{H, W\}\!\!\} + \frac{\kappa}{2} \left(    \partial_x x +  \partial_p p  \right) W + \frac{\kappa}{2} \left(\frac{1}{2} + \bar{n}_{\mathrm{th}}\right) \left( \partial^2_x  + \partial^2_p \right) W.
\end{split}
\end{align}
\Cref{eq:FP-Q} is the quantum version of the Fokker-Planck equation, with the Poisson bracket replaced by the Moyal bracket and a quantum diffusion term corresponding to the zero point spread.

Note that for the Hamiltonian corresponding to \cref{eq:H-eff}, the solution for $W$ from \cref{eq:FP-Q} will not yield the Boltzmann distribution in steady state, which perhaps is not surprising for an out-of-equilibrium driven problem \cite{dykman2012}.

\bibliography{bib_supp}
\end{widetext}